\documentclass[11pt,cls,onecolumn]{IEEEtran}

\usepackage{epsfig, amssymb}
\usepackage[cmex10]{amsmath}
\usepackage{subfigure}

\begin{document}

\title{Cognitive Networks Achieve Throughput Scaling of a Homogeneous Network}

\author{Sang-Woon Jeon, \emph{Student Member}, \emph{IEEE}, Natasha Devroye, Mai vu, \emph{Member}, \emph{IEEE}\\
Sae-Young Chung, \emph{Senior Member}, \emph{IEEE}, and Vahid Tarokh \emph{Fellow}, \emph{IEEE}\\
\thanks{S.-W. Jeon and S.-Y. Chung are with the School of EECS, KAIST, Daejeon, Korea (e-mail: swjeon@kaist.ac.kr; sychung@ee.kaist.ac.kr).}
\thanks{N. Devroye is with the Dept. of Electrical and Computer Engineering, University of Illinois at Chicago, Chicago, IL 60607 USA (e-mail: devroye@uic.edu).}
\thanks{M. Vu is with the Department of Electrical and Computer Engineering, McGill University, Montreal, Canada (e-mail: mai.h.vu@mcgill.ca).}
\thanks{V. Tarohk is with the School of Engineering and Applied Sciences, Harvard University, Cambridge, MA 02138 USA (e-mail: vahid@seas.harvard.edu).}
\thanks{The material in this paper was presented in part at the 7th International Symposium on Modeling and Optimization in Mobile, Ad Hoc, and Wireless Networks (WiOpt), Seoul, Korea, June 2009.}
 }

\maketitle


\newtheorem{definition}{Definition}
\newtheorem{theorem}{Theorem}
\newtheorem{lemma}{Lemma}
\newtheorem{example}{Example}
\newtheorem{corollary}{Corollary}
\newtheorem{proposition}{Proposition}
\newtheorem{conjecture}{Conjecture}
\newtheorem{remark}{Remark}

\def \diag{\operatornamewithlimits{diag}}
\def \min{\operatornamewithlimits{min}}
\def \max{\operatornamewithlimits{max}}
\def \log{\operatorname{log}}
\def \max{\operatorname{max}}
\def \rank{\operatorname{rank}}
\def \out{\operatorname{out}}
\def \exp{\operatorname{exp}}
\def \arg{\operatorname{arg}}
\def \E{\operatorname{E}}
\def \tr{\operatorname{tr}}
\def \SNR{\operatorname{SNR}}
\def \SINR{\operatorname{SINR}}
\def \dB{\operatorname{dB}}
\def \ln{\operatorname{ln}}
\def \th{\operatorname{th}}

\begin{abstract}
We study two distinct, but overlapping, networks that operate at the same
time, space, and frequency. The first network consists of $n$ randomly
distributed \emph{primary users}, which form either an ad hoc network, or
an infrastructure-supported ad hoc network with $l$ additional base
stations. The second network consists of $m$ randomly distributed, ad hoc
\emph{secondary users} or \emph{cognitive users}. The primary users have
priority access to the spectrum and do not need to change their
communication protocol in the presence of secondary users. The secondary
users, however, need to adjust their protocol based on knowledge about the
locations of the primary nodes to bring little loss to the primary
network's throughput. By introducing preservation regions around primary
receivers and avoidance regions around primary base stations, we propose
two modified multihop routing protocols for the cognitive users. Base on
percolation theory, we show that when the secondary network is denser than
the primary network, \emph{both} networks can \emph{simultaneously} achieve
the same throughput scaling law as a stand-alone network. Furthermore, the
primary network throughput is subject to only a vanishingly fractional loss.
Specifically, for the ad hoc and the infrastructure-supported primary
models, the primary network achieves sum throughputs of order $n^{1/2}$ and
$\max\{n^{1/2},l\}$, respectively. For both primary network models, for any
$\delta>0$, the secondary network can achieve sum throughput of order
$m^{1/2-\delta}$ with an arbitrarily small fraction of outage. Thus, almost
all secondary source-destination pairs can communicate at a rate of
order $m^{-1/2-\delta}$.
\end{abstract}


\section{Introduction}
In their pioneering work \cite{GuptaKumar:00}, Gupta and Kumar posed
and studied the limits of communication in ad hoc wireless
networks. Assuming $n$ nodes are uniformly distributed in a plane and
grouped into source-destination (S-D) pairs at random, they showed
that one  can achieve a sum throughput of
$S(n)=\Theta(\sqrt{n/\log n})$.  This is achieved using a  multihop
transmission scheme in which nodes transmit to one of the nodes in their neighboring cells, requiring full connectivity with at least one node per cell.
A trade-off between throughput and delay of fully-connected networks was studied in \cite{GamalMammenPrabhakarShah:04} and was extended in \cite{GamalMammen:06} to trade-offs between throughput, delay as well as energy.

The work in \cite{Olivier:06} has studied relay networks in which a single source transmits its data to the intended destination using the other nodes as relays.
Using  percolation theory \cite{Meester:96, Penrose:96}, they showed that a constant rate is achievable for a single S-D pair if we allow a small fraction of nodes to be disconnected.
This result can be applied to ad hoc networks having multiple S-D pairs and the work in \cite{Franceschetti:07} proposed an indirect multihop routing protocol based on such partial connectivity, that is all S-D pairs perform multihop transmissions based on this partially-connected sub-network.
They showed that the indirect multihop routing improves the achievable sum throughput as $S(n)=\Theta(\sqrt{n})$.

Information-theoretic outer bounds on throughput scaling laws of ad
hoc wireless networks were derived in \cite{AleksandarPramodSanjeev:04, XueXieKumar:05, LevequeTelatar:05, XieKumar:06}.
These bounds showed that the multihop routing using neighbor nodes is order-optimal in the power-limited and high attenuation regime.
Recently, a hierarchical
cooperation scheme was proposed in \cite{OzgurLevequeTse:06} and was
shown to achieve better throughput scaling than the multihop
strategy in the interference-limited or low attenuation regime, achieving a scaling very
close to their new outer bound. A more general hierarchical cooperation was proposed in \cite{Niesen:07}, which works for an arbitrary node distribution in which a minimum separation between nodes is guaranteed.

Recently \emph{hybrid network} models have been studied as well.
Hybrid networks are ad hoc networks in which the nodes' communication is aided by
additional infrastructures such as base stations (BSs). These are generally assumed to have high bandwidth connections to each other.
 In \cite{Dousse:02, Ganti:07} the connectivity of
hybrid networks has been analyzed.
In \cite{Agarwal:04, Liu:03,KulkarniViswanath:03, Zemlianov:05, LiuThiranTowsley:07} the throughput scaling of hybrid
networks has been studied. In order for a hybrid network's throughput
scaling to outperform that of a strictly ad hoc network, it was determined that the number of
BSs should be greater than a certain threshold \cite{Liu:03, Zemlianov:05}.

The existing literatures have focused on the throughput scaling of a
\emph{single} network. However, the necessity of extending and
expanding results to  capture \emph{multiple} overlapping networks
is becoming apparent. Recent measurements have shown that despite
increasing demands for bandwidth, much of the currently licensed
spectrum remains unused a surprisingly large portion of the time
\cite{FCC_measurements}. In the US, this has led the Federal
Communications Commission (FCC) to consider easing the regulations
towards \emph{secondary spectrum sharing} through their
\emph{Secondary Markets Initiative} \cite{SMI}. The essence of
secondary spectrum sharing involves having primary license
holders allow secondary license holders to access the
spectrum. Different types of spectrum sharing exist but most agree
that the primary users have a higher priority access to the spectrum,
while secondary users \emph{opportunistically} use it. These
secondary users often require greater sensing abilities and more
flexible and diverse  communication abilities than legacy primary
users. Secondary users are often assumed to be \emph{cognitive
radios}, or wireless devices which are able to transmit and
receive according to a variety of protocols and are also able to
sense and independently adapt to their environment
\cite{mitola_thesis}. These features allow them to behave in a more
``intelligent'' manner than current wireless devices.

In this paper, we consider \emph{cognitive networks}, which consist
of secondary, or cognitive, users who wish to transmit over the spectrum licensed to the primary users.
The single-user case in which a single primary and a single
cognitive S-D pairs share the spectrum  has been considered in the
literature, see for example \cite{Natasha:06, devroye_commmag,
Syed:07, jovicic:06} and the references therein. In \cite{Natasha:06}
the primary and cognitive S-D pairs are modeled as an
interference channel with asymmetric side-information.  In
\cite{Syed:07} the communication opportunities are modeled as a
two-switch channel.  Recently, a single-hop cognitive network was
considered in \cite{Mai:07}, where multiple secondary S-D pairs
transmit  in the presence of a single primary S-D pair.
It was shown that a linear scaling law of the single-hop secondary network is obtained
when its operation is constrained to guarantee a particular outage
constraint for the primary S-D pair.

We study a more general environment in which a \emph{ primary ad hoc network} and a \emph{cognitive ad hoc network} both share the same space, time and frequency dimensions.
Two types of primary networks are considered in this paper $\!$: an ad hoc primary network and an infrastructure-supported primary network.
For the ad hoc primary model, the primary network consists of $n$ nodes randomly distributed and grouped into S-D pairs at random.
For the infrastructure-supported primary model, additional $l$ BSs are regularly deployed and used  to support the primary transmissions.
In both cases, the cognitive network consists of $m$ secondary nodes distributed randomly and S-D pairs are again chosen randomly.
Our main assumptions are that (1) the primary network continues to operate as if no secondary network were present, (2) the secondary nodes know the locations of the primary nodes and (3) the secondary network is denser than the primary network.
Under these assumptions, we will illustrate routing protocols for the primary and secondary networks that result in the \emph{same throughput scaling} as if each were a single network.
Note that the constraint that the primary network does \emph{not} alter
its protocol because of the secondary network is what makes the
problem non-trivial. Indeed, if the primary network were to change
its protocol when the secondary network is present, a simple
time-sharing scheme is able to achieve the throughput scaling of homogeneous networks for both primary and secondary networks.

For the ad hoc primary model, we use a routing protocol that is a simple
modification of the nearest neighbor multihop schemes in
\cite{GuptaKumar:00,Franceschetti:07}.
For the infrastructure-supported primary model, we use a BS-based transmission similar to the scheme in \cite{Liu:03}. We propose novel routing protocols for the
secondary network under each primary network model. Our proposed protocols use multihop routing, in which the secondary routes \emph{avoid} passing too close to the
primary nodes, reducing the interference to them.
We show that the proposed protocols achieve the throughput scalings of homogeneous networks
\emph{simultaneously}. This implies that when a denser ``intelligent'' network is
layered on top of a sparser oblivious one, then both may achieve the same throughput scalings as if each were a single network. This result may be extended to more than two
networks, provided each layered network obeys the same three main
assumptions as in the two network case.

This paper is structured as follows. In Section \ref{sec:model} we
outline the system model: we first look at the network geometry,
co-existing primary and secondary ad hoc networks, then turn to the
information theoretic achievable rates before stating our
assumptions on the primary and secondary network behaviors. In
Section \ref{sec:adhoc} we outline the protocols used for the ad hoc primary model and prove that the claimed single network
throughput scalings may be achieved. We also prove the claimed single network
throughput scalings for the infrastructure-supported primary model in Section \ref{sec:infra}. We conclude in Section
\ref{sec:conclusion} and refer the proofs of the  lemmas to the Appendix.



\section{System Model}
\label{sec:model}

In order to study throughput scaling laws of ad hoc cognitive networks, we must define the underlying network models.
We first explain the two geometric models that will be considered in Sections \ref{sec:adhoc} and \ref{sec:infra}.
We then look at the transmission schemes, resulting achievable rates, and assumptions made about the primary and secondary networks.

Throughout this paper, we use $\mathbb{P}(E)$ to denote the probability of an event $E$ and we will be dealing with events which take
place with high probability (w.h.p.), or with probability 1 as the node
density tends to infinity\footnote{For simplicity, we use the notation w.h.p. in the paper to mean an event occurs with high probability as $n\rightarrow \infty$.}.

\subsection{Network Geometry}

\begin{figure}
  \begin{center}
  \scalebox{0.9}{\includegraphics{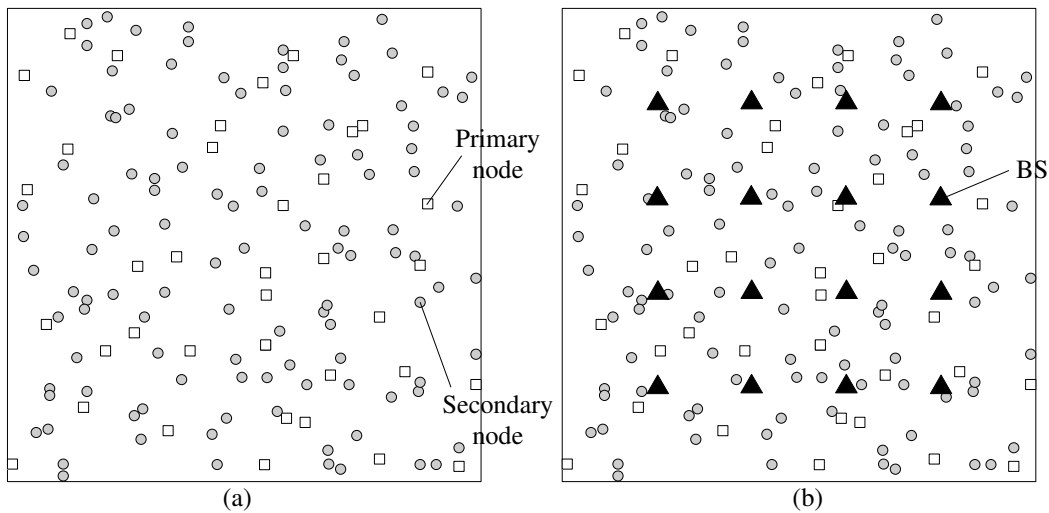}}
  \caption{We consider two network models. In (a), the primary nodes as well as the secondary nodes form distinct and co-existing ad hoc networks. This model is analyzed in Section \ref{sec:adhoc}. In (b), the primary nodes communicate with the help of BSs, while the secondary nodes still form an ad hoc network. This model is analyzed in Section \ref{sec:infra}.}
  \label{FIG:2models}
  \end{center}
\end{figure}

We consider a planar area in which a network of primary nodes and a network of secondary nodes co-exist. That is, the two networks share the same space, time, code, and frequency dimensions.
Two types of networks are considered as the primary network: an \emph{ad hoc network} and an
\emph{infrastructure-supported network}, while the secondary network is
always ad hoc. The two geometric models are illustrated in Fig.
 \ref{FIG:2models}.
As shown in Fig. \ref{FIG:2models}. (a), in the ad hoc primary model, nodes are distributed
according to a Poisson point process (p.p.p.) of density $n$ over a unit square, which are randomly grouped into primary S-D pairs. For the
secondary network, nodes are distributed according to a p.p.p. of density $m$ over the same unit square and are also randomly
grouped into secondary S-D pairs.

Our second model is the infrastructure-supported primary model, shown in Fig. \ref{FIG:2models}. (b).
There, primary nodes are still randomly distributed over the square according to a p.p.p. of density $n$, but these nodes are supported by additional
$l$ regularly spaced BSs (the number of BSs is equal to $l$, which is also the density of BSs).
The BSs' sole purpose is to relay data for the primary network, they
are neither sources nor destinations. We assume that the  BSs are connected to each other through wired lines of capacity large enough such that the BS-BS communication is not the limiting factor in the throughput scaling laws. Secondary nodes again form an ad hoc network with random S-D pairs, distributed according to a p.p.p. of density $m$.

The densities of the $n$ primary nodes, $m$ secondary nodes, and $l$ BSs are related according to
\begin{equation}
n=m^{\frac{1}{\beta}}=l^{\frac{1}{\gamma}},
\label{EQ:relations_l_m_n}
\end{equation}
where $\beta>1$ and $\gamma<1$.
We focus on the case where the density of the secondary nodes is higher than that of the primary nodes.
We also assume that the densities of both the primary nodes and secondary nodes are higher than that of the BSs, which is reasonable from a practical point of view.

The wireless propagation channel typically includes path loss with
distance, shadowing and fading effects. However, in this work we assume the channel gain depends only on the distance between a transmitter and its receiver, and ignore shadowing and fading.
Thus, the channel power gain $g(d)$, normalized by a constant, is given by
\begin{equation}
g(d)=d^{-\alpha},
\end{equation}
where $d$ denotes the distance between a transmitter (Tx) and its
receiver (Rx) and $\alpha>2$ denotes the path-loss exponent.

\subsection{Rates and Throughputs Achieved}

Each network operates based on slotted transmissions. We assume the duration
of each slot, and the coding scheme employed are such that one can achieve the additive white Gaussian noise (AWGN) channel capacity.
For a given signal to interference and noise ratio (SINR), this capacity is given by the well known formula $R=\log(1+\operatorname{SINR})$ bps/Hz assuming the additive interference is also white, Gaussian, and independent from the noise and signal.
We assume that primary slots and secondary slots have the same
duration and are synchronized with each other.
We further assume all the primary, secondary, and BS nodes are subject to a transmit power constraint $P$.

We now characterize the rates achieved by the primary and secondary transmit pairs.
Suppose that $N_p$ primary pairs and $N_s$ secondary
pairs communicate simultaneously. Before proceeding with a detailed description, let
us define the notations used in the paper, given by Table \ref{Table:simbols}.
Then, the $i$-th primary pair can communicate at a rate of
\begin{equation}
R_p^i=\log\left(1+\frac{P_p^ig\left(\|X^i_{p,\text{tx}}-X_{p,\text{rx}}^i\|\right)}{N_0+I_p^i+I_{sp}^i}\right),
\end{equation}
where $\|\cdot\|$ denotes the Euclidean norm of a vector. $I_p^i$
and $I_{sp}^i$ are given by
\begin{equation}
I_p^i=\sum_{k=1,k\neq
i}^{N_p}P_p^kg\left(\|X_{p,\text{tx}}^k-X_{p,\text{rx}}^i\|\right)
\end{equation}
and
\begin{equation}
I_{sp}^i=\sum_{k=1}^{N_s}P_s^kg\left(\|X_{s,\text{tx}}^k-X_{p,\text{rx}}^i\|\right).
\end{equation}
Similarly, the $j$-th secondary pair can communicate at a rate of
\begin{equation}
R_s^j=\log\left(1+\frac{P_s^jg\left(\|X_{s,\text{tx}}^j-X_{s,\text{rx}}^j\|\right)}{N_0+I_s^j+I_{ps}^j}\right),
\end{equation}
where $I_s^j$ and $I_{ps}^j$ are given by
\begin{equation}
I_s^j=\sum_{k=1,k\neq
j}^{N_s}P_s^kg\left(\|X_{s,\text{tx}}^k-X_{s,\text{rx}}^j\|\right)
\end{equation}
and
\begin{equation}
I_{ps}^j=\sum_{k=1}^{N_p}P_p^kg\left(\|X_{p,\text{tx}}^k-X_{s,\text{rx}}^j\|\right).
\end{equation}

Throughout the paper, the achievable per-node throughput of the
primary and secondary networks are defined as follows.
\begin{definition}
A throughput of $T_p(n)$ per primary node is said to be achievable w.h.p. if all primary sources can transmit at a rate of $T_p(n)$ (bps/Hz) to their primary destinations w.h.p. in the presence of the secondary network.
\end{definition}
\begin{definition}
Let $\epsilon_s(m)>0$ denote an outage probability of the secondary network, which may vary as a function of $m$.
A throughput of $T_s(m)$ per secondary node is said to be $\epsilon_s(m)$-achievable w.h.p. if at least $1-\epsilon_s(m)$ fraction of secondary sources can transmit at a rate of $T_s(m)$ (bps/Hz) to their secondary destinations w.h.p. in the presence of the primary network.
\end{definition}

For both ad hoc and infrastructure-supported primary models, we will propose secondary routing schemes that make $\epsilon_s(m)\to 0$ as $m\to\infty$\footnote{In this paper, $m\rightarrow\infty$ is equivalent to $n\rightarrow\infty$ since $m=n^{\beta}$.}.
Thus, although we allow a fraction of secondary S-D pairs to be in outage, for sufficiently large $m$, almost all secondary S-D pairs will be served at a rate of $T_s(m)$.
Let us define $S_p(n)$ as the sum throughput of the primary network,
or  $T_p(n)$ times the number of primary S-D pairs\footnote{We note that in general $S_p(n)\neq \frac{n}{2}T_p(n)$ since the nodes are thrown at random according to a p.p.p. of density $n$. The actual number of nodes in the network will vary in a particular realization.}.
Similarly, we define $S_s(m)$ as the sum throughput of the secondary
network, or  $T_s(m)$ times the number of \emph{served}
secondary S-D pairs at a rate of $T_s(m)$. While $T_p(n)$ and $S_p(n)$ represent the per-node and sum throughputs of the primary network \emph{in the presence of the secondary network}, we use
the notations $T_{\operatorname{alone}}(n)$ and $S_{\operatorname{alone}}(n)$ to denote the per-node and sum throughputs of the primary network \emph{in the
absence of the secondary network}, respectively.

\subsection{Primary and Secondary User Behaviors}

As primary and secondary nodes must share the spectrum, the rules or assumptions made about this co-existence are of critical importance to the resulting achievable throughputs and scaling laws.
Primary networks may be thought of as existing
communication systems that operate in licensed bands. These primary users are the license holders, and thus have higher priority access to the spectrum than secondary users. Thus, our first key assumption is that \emph{the primary network does not have to change its protocol due to the
secondary network.} In other words, all primary S-D pairs communicate
with each other as intended, regardless of the secondary network.
The secondary network, which is opportunistic in nature,  is responsible for reducing its interference to
the primary network to an ``acceptable level'', while maximizing its own throughput $T_s(m)$. This acceptable level may be defined to be one that does not degrade the throughput scaling of the primary network.
More strictly, the secondary network should satisfy w.h.p.
\begin{equation}
\frac{T_p(n)}{T_{\operatorname{alone}}(n)}\geq 1-\delta_{\operatorname{loss}}
\label{EQ:fraction_loss}
\end{equation}
during its transmission, where $\delta_{\operatorname{loss}}\in(0,1)$ is the maximum allowable fraction of throughput loss for the primary network.
Notice that the above condition guarantees $T_p(n)=\Theta\left(T_{\operatorname{alone}}(n)\right)$.
The secondary network  may ensure (\ref{EQ:fraction_loss}) by adjusting its protocol based on information
about the primary network.
Thus, our second key assumption is that \emph{the secondary network knows the locations of all primary
nodes.}
Since the secondary network is denser than the primary network, each secondary node can measure the interference power from its
adjacent primary node and send it to a coordinator node.
Based on these measured values, the secondary network can establish the locations of the primary nodes.

\section{Ad Hoc Primary Network}
\label{sec:adhoc}

We first consider the  throughput scaling laws when both the primary and secondary networks are ad hoc in nature.
Since the primary network needs not change its transmission scheme due to the presence of the secondary network,  we assume it transmits according to the direct multihop routing similar to those in \cite{GuptaKumar:00} and \cite{GamalMammenPrabhakarShah:04}.
We also consider the indirect multihop routing proposed in \cite{Franceschetti:07} as a primary protocol.
Of greater interest is how the secondary nodes will transmit such that the primary network remains unaffected in terms of throughput scaling.

\subsection{Main Results}

The main results of this section describe achievable throughput scaling laws of the primary and secondary networks. We simply  state these results here and derive them in the remainder of this section.

{\em
Suppose the ad hoc primary model.
For any $\delta_{\operatorname{loss}}\in(0,1)$, the primary network can achieve the following per-node and sum throughputs w.h.p.:
\begin{equation}
T_p(n)=(1-\delta_{\operatorname{loss}})T_{\operatorname{alone}}(n), \hspace{2cm}S_p(n)=(1-\delta_{\operatorname{loss}})S_{\operatorname{alone}}(n),
\end{equation}
where
\begin{equation}
T_{\operatorname{alone}}(n)=\begin{cases}\Theta\left(\frac{1}{\sqrt{n\log n}}\right)&\mbox{for direct multihop routing}\\
\Theta\left(\frac{1}{\sqrt{n}}\right)&\mbox{for indirect multihop routing}\\
\end{cases}
\end{equation}
and $S_{\operatorname{alone}}(n)=\Theta\left(nT_{\operatorname{alone}}(n)\right)$.
The following per-node and sum throughputs are $\epsilon_s(m)$-achievable w.h.p. for the secondary network:
\begin{equation}
T_s(m)=\Theta\left(\frac{1}{\sqrt{m\log m}}\right),\hspace{2cm}S_s(m)=\Theta\left(\sqrt{\frac{m}{\log m}}\right),
\end{equation}
where $\epsilon_s(m)=\operatorname{O}\left(\frac{\log m}{m^{1-1/\beta}}+\frac{\sqrt{\log m}}{m^{3/2-2/\beta}}\right)$, which converges to zero as $m\to \infty$.
}

\vspace{0.1 in}
This result is of particular interest as it shows that not only can the primary network operate at the same scaling law as when the secondary network does not exist, but the secondary network can also achieve, with an arbitrarily small fraction of outage, the exact same scaling law obtained by the direct multihop routing as when the primary network does not exist.
Thus almost all secondary S-D pairs can communicate at a rate of $T_s(m)$ in the limit of large $m$.
In essence, whether the indirect multihop or the direct multihop is adopted as a primary protocol, the secondary network can achieve the sum throughput of $\Theta(\sqrt{m/\log m})$ w.h.p. while preserving $1-\delta_{\operatorname{loss}}$ fraction of the primary network's stand-alone throughput.


In the remainder of this section, we first outline the operation of the primary network and then focus on the design of a secondary network protocol under the given primary protocol.
We analyze achievable throughputs of the primary and secondary networks, which will determine the  throughput scaling of both co-existing networks. Throughout this work, we place the proofs of more technical lemmas and theorems in the Appendix and outline the main proofs in the text.

\subsection{Network Protocols}
We assume the primary network communicates according to the direct multihop routing protocol.
The indirect multihop routing will be explained in Section \ref{subsec:indirect_routing}, which can be extended from the results of the direct routing.
The challenge is thus to prove that the secondary nodes can exchange information in such a way that satisfies $T_p(n)\geq (1-\delta_{\operatorname{loss}}) T_{\operatorname{alone}}(n)$ w.h.p..
We first outline a primary network protocol, and then design a secondary network protocol which operates in the presence of the primary network.

\subsubsection{Primary network protocol}

We assume that the primary network delivers data using the direct
multihop routing, in a manner similar to \cite{GuptaKumar:00} and
\cite{GamalMammenPrabhakarShah:04}.
The basic multihop protocol is as follows:
\begin{itemize}
\item Divide the unit area into square cells of area $a$.
\item A $9$- time division multiple access (TDMA) scheme is used,
  in which each cell is activated during one out of $9$ slots.
\item Define the horizontal data path (HDP) and the vertical data path
  (VDP) of a S-D pair as the horizontal line and the vertical
  line connecting a source to its destination, respectively.
  Each source transmits data to its destination by first hopping to
  the adjacent cells on its HDP and then on its VDP.
\item When a cell becomes active, it delivers its
  traffic. Specifically, a Tx node in the active cell transmits a
  packet to a node in an adjacent cell (or in the same
  cell). A simple round-robin scheme is used for all Tx nodes in the
  same cell.
\item At each transmission, a Tx node transmits with power $P{d}^{\alpha}$, where $d$ denotes the distance between the Tx and its Rx.
\end{itemize}

This protocol requires full connectivity, meaning that each cell should have at least one node.
Let $a_p$ denote the area of a primary cell. The following lemma indicates how to determine $a_p$ satisfying this requirement.

\begin{lemma} \label{THM:node_number}
 The following facts hold. \\
  (a) The number of primary nodes in a unit area is within
  $\left((1-\epsilon)n, (1+\epsilon)n\right)$ w.h.p., where $\epsilon>0$ is
  an arbitrarily small constant. \\
  (b) Suppose $a_p=\frac{2\log n}{n}$. Then, each primary cell has at least one primary node w.h.p..
\end{lemma}
\begin{proof}
The proof is in the Appendix.
\end{proof}

Based on Lemma \ref{THM:node_number}, we set $a_p=\frac{2\log n}{n}$.
Under the given primary protocol, $T_{\operatorname{alone}}(n)=\Theta(1/\sqrt{n \log n})$ and $S_{\operatorname{alone}}(n)=\Theta(\sqrt{n/\log n})$ are achievable w.h.p. when the secondary network is absent or silent.

Results similar to Lemma \ref{THM:node_number} can be found in \cite{GuptaKumar:00} and \cite{GamalMammenPrabhakarShah:04}, where their proposed schemes also achieve the same $T_{\operatorname{alone}}(n)$ and $S_{\operatorname{alone}}(n)$.
Note that the Gupta-Kumar's model \cite{GuptaKumar:00}, \cite{GamalMammenPrabhakarShah:04} assumes uniformly distributed nodes in the network and a constant rate between Tx and Rx if SINR is higher than a certain level.
Although we assume that the network is constructed according to a p.p.p. (rather than uniform) and that the information-theoretic rate $\log(1+\operatorname{SINR})$ is achievable (rather than a constant rate), the above primary network protocol provides the same throughput scaling as that under the Gupta-Kumar's model.

\subsubsection{Secondary network protocol}
Since the secondary nodes know the primary nodes' locations, an
intuitive idea is to have the secondary network operate in a
multihop fashion in which they circumvent each primary node in order
to reduce the effect of secondary transmissions to the  primary
nodes.
In \cite{Fang:04, Subramanian:07} a network with holes is considered and geographic forwarding algorithms that establish routing paths around holes are proposed.

\begin{figure}
  \begin{center}
  \scalebox{0.9}{\includegraphics{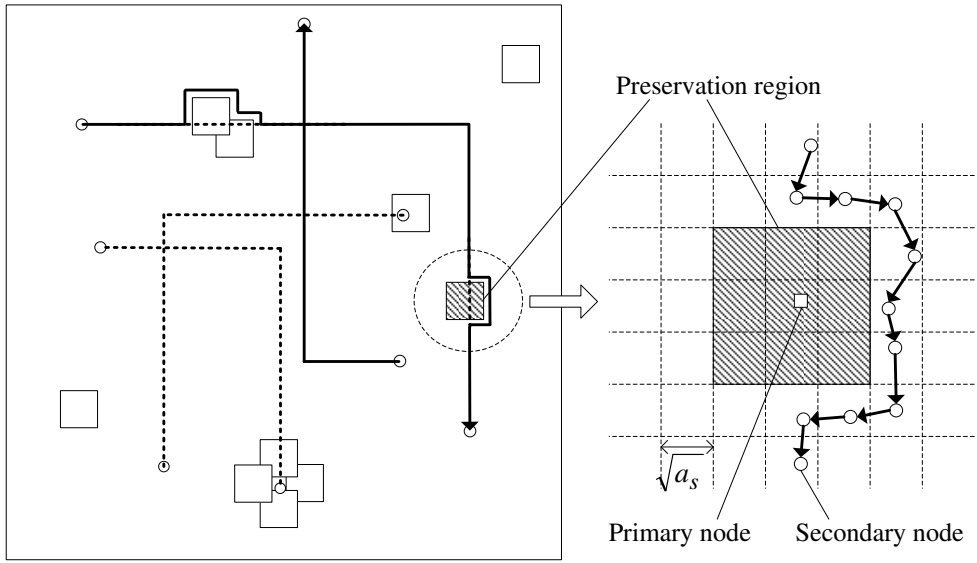}}
  \caption{Secondary data paths for the ad hoc primary model: a secondary S-D pair goes around if it is blocked by a preservation region. If a source (or its destination) is in a preservation region or its data path is disconnected by preservation regions, the corresponding S-D pair is not served.}
  \label{FIG:secodary_routing_adhoc}
  \end{center}
\end{figure}

Around each primary node we define its \emph{preservation
region:} a square containing $9$ secondary cells, with the primary
node at the center cell. The secondary nodes, when determining their routing paths,  need to avoid these preservation regions:
Our protocol for the secondary ad hoc network is the same as the basic multihop protocol except that
\begin{itemize}
\item The secondary cell size is $a_s=\frac{2\log m}{m}$.
\item At each transmission a secondary node transmits its packet \emph{three} times repeatedly (rather than once) using three slots.
\item The secondary paths avoid the preservation
regions (see Fig. \ref{FIG:secodary_routing_adhoc}). That is, if the HDP or VDP of a
secondary S-D pair is blocked by a preservation region, this data
path circumvents the preservation region by  using its adjacent cells.
If a secondary source (or its destination) belongs to
preservation regions or its data path is disconnected by preservation regions, the
corresponding S-D pair is not served.
\item At each transmission, a Tx node transmits with power $\delta_{P}P{d}^{\alpha}$, where $d$ denotes the distance between the Tx and its Rx and $\delta_P\in(0,1)$.
\end{itemize}

Since $a_s$ converges to zero as $m\to \infty$, there exists $m_0$ such that the power constraint is satisfied for any $\delta_P\in(0,1)$ if $m\geq m_0$.
We will show in Lemma \ref{THM:primary_rate_adhoc1} that adjusting $\delta_P$ induces a trade-off between the rates of the primary and secondary networks while the scaling laws of both networks are unchanged, which allows the condition (\ref{EQ:fraction_loss}) to be meet.

Unlike the primary protocol, each secondary cell transmits a
secondary packet three times repeatedly when it is activated.
As we will show later,  the repeated secondary transmissions can
guarantee the secondary receivers a certain minimum distance from all
primary interferers for at least one packet, thus guaranteeing the
secondary network a non-trivial rate.
Therefore, the duration of the secondary $9$-TDMA scheme is three
times longer than that of the primary $9$-TDMA.
The main difference between this scheme and previous multihop routing schemes is that the
secondary multihop paths must circumvent the preservation regions
and that a portion of secondary S-D pairs is not served.
But this portion will be negligible as $m\to\infty$.
By re-routing the secondary nodes'
transmission around the primary nodes' preservation regions, we can
guarantee the primary nodes a non-trivial rate.


Similar to Lemma \ref{THM:node_number}, we can also prove that the total
number of secondary nodes is within $((1-\epsilon)m, (1+\epsilon)m)$
w.h.p. and that each secondary cell has at least one secondary node w.h.p..

\subsection{Throughput Analysis and its Asymptotic Behavior}
In this subsection, we analyze the per-node and sum throughputs of each network under the given protocols and derive throughput scaling laws with respect to the node densities.

\subsubsection{Primary network throughputs}


Let us consider the primary network in
the presence of the secondary network. We first show that each primary cell can sustain a constant aggregate rate (Lemma \ref{THM:primary_rate_adhoc1}), which may be used in conjunction with the number of data paths each primary cell must transmit (Lemma \ref{THM:primary_traffic_adhoc}) to obtain the per-node and sum throughputs in Theorem \ref{THM:primary_throughput_adhoc}.

Let $R_p(n)$ and $R_{\operatorname{alone}}(n)$ denote the achievable aggregate rate of each primary cell in the presence and in the absence of the secondary network, respectively.
We define
\begin{equation}
I\triangleq P2^{\alpha/2+3}\sum_{t=1}^{\infty}t(3t-2)^{-\alpha}
\label{EQ:def_I}
\end{equation}
having a finite value for $\alpha>2$, which will be used to derive an upper bound on the interference power of the ad hoc primary and secondary networks.
Then the following lemma holds.

\begin{lemma} \label{THM:primary_rate_adhoc1}
Suppose the ad hoc primary model.
If $\delta_P\in(0,\min\{\delta_{P,\operatorname{max}},1\})$, then
\begin{equation}
\lim_{n\to\infty}\frac{R_p(n)}{R_{\operatorname{alone}}(n)}\geq 1-\delta_{\operatorname{loss}},
\end{equation}
where $\delta_{P,\max}=\big(\frac{1}{(1+\frac{P}{N_0})^{1-\delta_{\operatorname{loss}}}-1}-\frac{N_0}{P}\big)\frac{P}{I}$ and $I$ is given by (\ref{EQ:def_I}).
Moreover, $R_{\operatorname{alone}}(n)$ is lower bounded by $K_p$, where $K_p=\frac{1}{9}\log\left(1+\frac{P}{N_0+I}\right)$ is a constant independent
of $n$.
\end{lemma}

\begin{proof}
The proof is in the Appendix.
\end{proof}

The essence of the proof of Lemma \ref{THM:primary_rate_adhoc1} lies in showing that the secondary nodes, even as $m\to\infty$,  do not cause the aggregate rate of each primary cell to decay with $m$.
This is done by introducing the preservation regions, which guarantee the minimum distance of $\sqrt{a_s}$ from all secondary Txs to the primary Rxs.
This Lemma will be used to show that (\ref{EQ:fraction_loss}) can be satisfied w.h.p. if $\delta_P\in(0,\min\{\delta_{P,\operatorname{max}},1\})$ in Theorem \ref{THM:primary_throughput_adhoc}.

\begin{figure}
  \begin{center}
  \scalebox{0.9}{\includegraphics{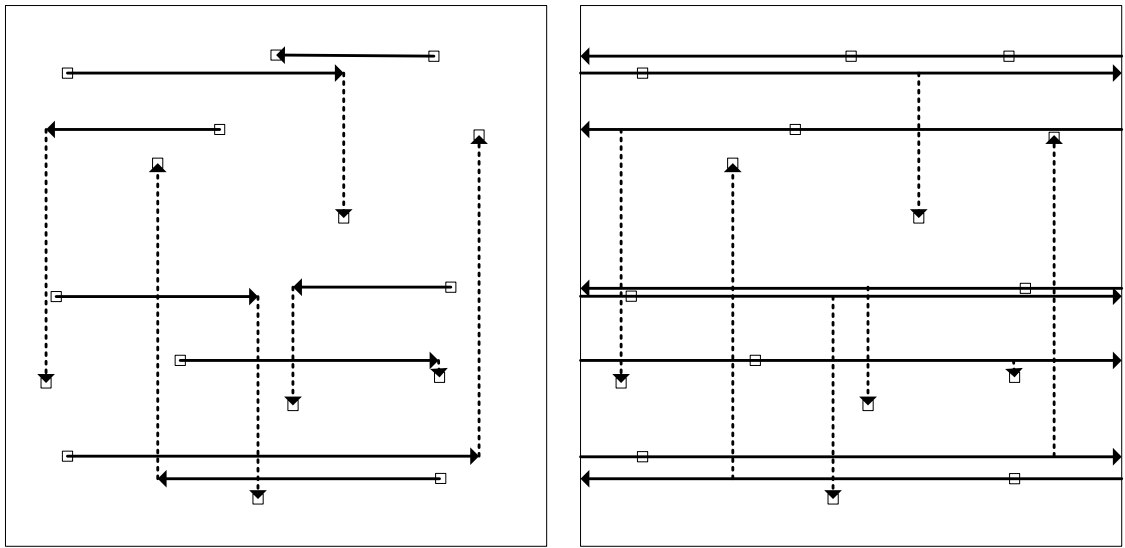}}
  \caption{Examples of original HDPs (left) and their extended HDPs (right) of the primary S-D pairs for the ad hoc primary model.}
  \label{FIG:routing_analysis_adhoc}
  \end{center}
\end{figure}

The next lemma determines the number of data paths that each cell should carry.
To obtain an upper bound, we extend each HDP to the entire horizontal line and all cells through which this horizontal line passes should deliver the corresponding data of HDP (see Fig. \ref{FIG:routing_analysis_adhoc}).
Similarly, we extend each VDP to the entire vertical line.
We define this entire horizontal and vertical line as an \emph{extended} HDP and an \emph{extended} VDP, respectively.
Throughout the rest of the paper, our analysis deals with extended HDPs and VDPs instead of original HDPs and VDPs.
Since we are adding hops to our routing scheme, the extended traffic gives us a lower bound on the achievable throughput.

\begin{lemma} \label{THM:primary_traffic_adhoc}
Under the ad hoc primary model, each primary cell needs to carry at most $4\sqrt{2n\log n}$ data paths w.h.p..
\end{lemma}
\begin{proof}
The proof is in the Appendix.
\end{proof}

Lemma \ref{THM:primary_traffic_adhoc} shows how the number of data
paths varies with the node density $n$. Lemmas
\ref{THM:node_number}-\ref{THM:primary_traffic_adhoc} may be used to
prove the main theorem, stated next.
\begin{theorem} \label{THM:primary_throughput_adhoc}
Suppose the ad hoc primary model.
For any $\delta_{\operatorname{loss}}\in(0,1)$, by setting $\delta_P\in(0, \min\{\delta_{P,\operatorname{max}},1\})$, the primary network can achieve $T_p(n)=(1-\delta_{\operatorname{loss}})T_{\operatorname{alone}}(n)$
and $S_p(n)=(1-\delta_{\operatorname{loss}})S_{\operatorname{alone}}(n)$ w.h.p., where
\begin{equation}
T_{\operatorname{alone}}(n)=\frac{K_p}{4\sqrt{2}}\frac{1}{\sqrt{n\log n}}
\end{equation}
and
\begin{equation}
S_{\operatorname{alone}}(n)=\frac{K_p(1-\epsilon)}{8\sqrt{2}}\sqrt{\frac{n}{\log n}}.
\end{equation}
The definitions of $\delta_{P,\operatorname{max}}$ and $K_p$ are given in Lemma \ref{THM:primary_rate_adhoc1}.
\end{theorem}
\begin{proof}
First consider the stand-alone throughput of the primary network.
Since each primary cell can sustain a rate of $K_p$ (Lemma
\ref{THM:primary_rate_adhoc1}), each primary S-D pair can achieve a
rate of at least $K_p$ divided by the maximum number of data paths per
primary cell. The number of data paths is upper bounded by
$4\sqrt{2n\log n}$ w.h.p. (Lemma \ref{THM:primary_traffic_adhoc}).
Therefore, $T_{\operatorname{alone}}(n)$ is lower bounded by
$\frac{K_p}{4\sqrt{2n\log n}}$ w.h.p..
Now the whole network contains at least $(1-\epsilon)\frac{n}{2}$
primary S-D pairs w.h.p. (Lemma \ref{THM:node_number}). Therefore, $S_{\operatorname{alone}}(n)$ is lower bounded by
$(1-\epsilon)\frac{n}{2}T_{\operatorname{alone}}(n)$ w.h.p..

Finally Lemma \ref{THM:primary_rate_adhoc1} shows that, for any $\delta_{\operatorname{loss}}\in(0,1)$, if we set $\delta_P\in(0,\min\{\delta_{P,\operatorname{max}},1\})$, then $R_p(n)= (1-\delta_{\operatorname{loss}})R_{\operatorname{alone}}(n)$ is achievable in the limit of large $n$.
Since the number of primary data paths carried by each primary cell and the total number of primary S-D pairs in the network holds regardless of the existence of the secondary network, $T_p(n)=(1-\delta_{\operatorname{loss}})T_{\operatorname{alone}}(n)$ and $S_p(n)=(1-\delta_{\operatorname{loss}})S_{\operatorname{alone}}(n)$ are also achievable w.h.p., which completes the proof.
\end{proof}

\subsubsection{Secondary network throughputs}
Let us now consider the per-node throughput of the secondary network in
the presence of the primary network.
The main difference between the primary and secondary transmission schemes arises from the presence of the preservation regions. Recall that the secondary nodes wish to transmit according to a multihop protocol, but their path may be blocked by a preservation region. In this case, they must circumvent the preservation region, or possibly the \emph{cluster} of primary preservation regions\footnote{Since the primary nodes are distributed according to a p.p.p., clustering of preservation regions may occur.}.
However, as we will see later circumventing these preservation regions (clusters) does not degrade the secondary network's throughput scaling due to the relative primary and secondary node densities: the secondary nodes increase at the rate $m=n^\beta$ and $\beta>1$.
Thus, intuitively, as the density $n$ of the primary nodes increases, the area of
each preservation region (which equals 9 secondary cells) decreases
faster than the increase rate of the primary node density (and thus
number of preservation regions). These clusters of preservation
regions remain bounded in size, although their number diverges as $n \rightarrow \infty$. This can be
obtained using  percolation theory \cite{Meester:96}.

Let us introduce a Poisson Boolean model $(X,\rho,\lambda)$ on $\mathbb{R}^d$.
The points $X_1,X_2,\cdots$ are distributed according to a p.p.p. of density $\lambda$ and each point $X_i$ is the center of a closed ball with radius $\rho_i$.
Notice that $\rho_i$'s are random variables independent of each other and independent of $X$, whose distributions are identical to that of $\rho$.
The \emph{occupied} region is the region that is covered by at least one ball and the \emph{vacant} region is the complement of the occupied region.
Note that the occupied (or vacant) region may consists of several occupied (vacant) components that are disjointed with each other.
Then the following theorem holds.

\begin{theorem}[Meester and Roy]  \label{THM:percolation}
For a Poisson Boolean model $(X,\rho,\lambda)$ on $\mathbb{R}^d$, for $d\geq2$, if $\mathbb{E}(\rho^{2d-1})< \infty$,
then there exists $\lambda_0>0$ such that
for all $0<\lambda<\lambda_0$,
\begin{equation}
\mathbb{P}(\mbox{number of balls in any occupied component is finite})=1.
\label{EQ:percolation}
\end{equation}
\end{theorem}
\begin{proof}
We refer readers to the proof of Theorem 3.3 in \cite{Meester:96}.
\end{proof}

By scaling the size of the above Poisson Boolean model and setting $\rho$ as a deterministic value, we apply Theorem \ref{THM:percolation} to our network model.

\begin{figure}
  \begin{center}
  \scalebox{1.2}{\includegraphics{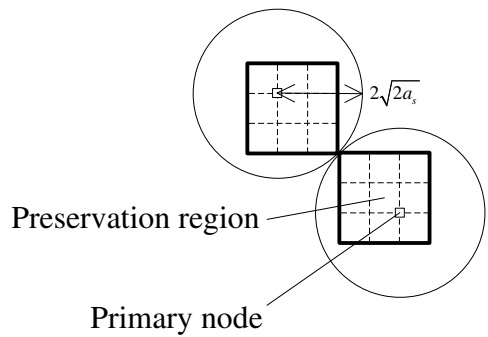}}
  \caption{Minimum distance between any two preservation regions such that the corresponding balls are not overlapping.}
  \label{FIG:percolation}
  \end{center}
\end{figure}

\begin{corollary} \label{THM:clusters_of_preservation_regions}
Any cluster of preservation regions
has at most $N_c$ preservation regions w.h.p., where $N_c>0$ is an integer independent of $n$.
\end{corollary}
\begin{proof}
Let us consider a Poisson Boolean model $(X,\rho=1,\lambda=8n a_s)$ on $\mathbb{R}^2$.
All balls in this model have deterministic radii of $1$ and the density of the underlining p.p.p. is a function of $n$ decreasing to zero as $n\to\infty$.
Thus, $\mathbb{E}(\rho^3)=1< \infty$ and there exists $n_0>0$ such that $\lambda<\lambda_0$ for all $n\geq n_0$.
As a consequence, (\ref{EQ:percolation}) holds for all $n\geq n_0$.
Since this result holds on $\mathbb{R}^2$, the same result still holds if we focus on the area of $\left[0,\frac{1}{2\sqrt{2a_s}}\right]^2$ instead of $\mathbb{R}^2$.
Moreover, two Poisson Boolean models $(X,\rho=1,\lambda=8na_s)$ on $\left[0,\frac{1}{2\sqrt{2a_s}}\right]^2$ and $(X',\rho'=2\sqrt{2a_s},\lambda'=n)$ on $\left[0,1\right]^2$ show the same percolation result (see Proposition 2.6.2 in \cite{Franceschetti2:07}).
Therefore, under the Poisson Boolean model $(X',\rho'=2\sqrt{2a_s},\lambda'=n)$ on $\left[0,1\right]^2$, the number of balls in any occupied component is upper bounded by $N_c$ w.h.p., where $N_c>0$ is an integer independent of $n$.

In the case of $(X',\rho'=2\sqrt{2a_s},\lambda'=n)$ on $\left[0,1\right]^2$, the underlining p.p.p. is the same as that of the primary network and each ball contains the corresponding preservation region shown in Fig. \ref{FIG:percolation}.
Thus preservation regions cannot form a cluster if the corresponding balls do not form an occupied component, meaning the number of preservation regions in any cluster is also upper bounded by $N_c$ w.h.p., which completes the proof.
\end{proof}

This corollary is needed to ensure that the secondary network remains
connected,  to bound the number of data paths that pass through
secondary cells, and to prove the next lemma.  As mentioned earlier,
whenever a secondary source or destination lies within a primary preservation
region or there is no possible data path, this pair is not
served. The next lemma shows that the fraction of these unserved
secondary S-D pairs is arbitrarily small w.h.p..

\begin{lemma}\label{THM:number_of_served_SD_adhoc}
Under the ad hoc primary model, the fraction of unserved secondary S-D pairs is upper bounded by $\epsilon_{s,1}(m)=\Theta(\frac{\log m}{m^{1-1/\beta}})$ w.h.p.,
which converges to zero as $m\rightarrow \infty$.
\end{lemma}
\begin{proof}
The proof is in the Appendix.
\end{proof}

Next, Lemma \ref{THM:secondary_rate_adhoc} shows that, in the presence of the
primary network, each secondary cell may sustain a constant aggregate
rate.

\begin{lemma} \label{THM:secondary_rate_adhoc}
Under the ad hoc primary model, each secondary
cell can sustain traffic at a rate of $K_s$, where $K_s=\frac{1}{27}\log\left(1+\frac{\delta_P P}{N_0+(1+\delta_P)I+2^{3\alpha/2}P}\right)$ is a constant
independent of $m$ and $I$ is given by (\ref{EQ:def_I}).
\end{lemma}
\begin{proof}
The proof is in the Appendix.
\end{proof}

The main challenge in proving Lemma
\ref{THM:secondary_rate_adhoc} is the presence of the primary
Txs. Since the primary node density is smaller than the secondary node
density, the primary cells are relatively further away from each
other, thus requiring higher power to communicate. Although the
relatively higher power could be a potential problem because the
secondary nodes repeat their transmissions for three slots, the interfering primary transmission occurs at a
certain minimum distance away from the secondary Rx on one
of these slots. Although the
actual rate of the secondary network is reduced by a factor of three,
this allows us to bound the interference of the more powerful primary
nodes, without changing the scaling laws.
From Lemma \ref{THM:primary_rate_adhoc1}, the value of $\delta_P$, which is a normalized transmit power of the secondary Txs, should be smaller than $\min\{\delta_{P,\operatorname{max}},1\}$ in order to satisfy (\ref{EQ:fraction_loss}).
We also notice that the range of $\delta_P$ does not affect the throughput scalings of the secondary network.

Let us define the secondary cells that border the preservation regions as \emph{loaded} cells and the other cells as \emph{regular} cells.
The loaded cells will be required to carry not only their own traffic, but also re-routed traffic around the preservation regions and, as a result, could deliver more data than the regular cells.
The next lemma bounds the number of data paths that each regular cell and each loaded cell must transport.
As the number of data paths each cell could carry was essentially the limiting factor in the sum throughput of the primary network, the following lemma is of crucial importance for the secondary sum throughput scaling law.

\begin{lemma} \label{THM:secondary_traffic_adhoc}
Under the ad hoc primary model, each regular secondary cell needs to carry at
most $4\sqrt{2m\log m}$ data paths and each loaded secondary cell carries at most $4(6N_c+1)\sqrt{2m\log m}$ data paths w.h.p., where $N_c$ is given in Corollary \ref{THM:clusters_of_preservation_regions}.
\end{lemma}
\begin{proof}
The proof is in the Appendix.
\end{proof}

As it will be shown later, for $1<\beta\leq4/3$ the loaded cells are the
bottleneck of the overall throughput. But even in this case, only a
constant fraction of throughput degradation occurs, which does not
affect the throughput scaling. For $\beta > 4/3$, since the secondary
network is much denser than the primary network, the fraction of
secondary data paths needing to be re-routed diminishes to zero as
the node densities increase. Thus in the limit, almost all secondary cells behave as regular
cells.

Finally, we can use the previous corollary and lemmas to obtain the
per-node and sum throughputs of the secondary network in the
following theorem.

\begin{theorem} \label{THM:secondary_throughput_adhoc}
Suppose the ad hoc primary model.
For any $\delta_{\operatorname{loss}}\in(0,1)$, by setting $\delta_P\in(0,\min\{\delta_{P,\operatorname{max}},1\})$,
the following per-node and sum throughputs are $\epsilon_s(m)$-achievable w.h.p. for the secondary network:
\begin{equation}
T_s(m)=\begin{cases} \frac{K_s}{4\sqrt{2}}\frac{1}{\sqrt{m\log m}}&\text{if $\beta>\frac{4}{3}$}\\
\frac{K_s}{4\sqrt{2}(6N_c+1)}\frac{1}{\sqrt{m\log m}}&\text{if $1<\beta\leq\frac{4}{3}$}\end{cases}
\end{equation}
and
\begin{equation}
S_s(m)=\begin{cases} \frac{K_s(1-\epsilon)(1-\epsilon_s(m))}{8\sqrt{2}}\sqrt{\frac{m}{\log m}}&\text{if $\beta>\frac{4}{3}$}\\
\frac{K_s(1-\epsilon)(1-\epsilon_s(m))}{8\sqrt{2}(6N_c+1)}\sqrt{\frac{m}{\log m}}&\text{if $1<\beta\leq\frac{4}{3}$},\end{cases}
\end{equation}
\end{theorem}
where $\epsilon_s(m)=\operatorname{O}\left(\frac{\log m}{m^{1-1/\beta}}+\frac{\sqrt{\log m}}{m^{3/2-2/\beta}}\right)$, which converges to zero as $m\to\infty$.
The definitions of $\delta_{P,\operatorname{max}}$, $K_s$, and $N_c$ are given in Lemma \ref{THM:primary_rate_adhoc1}, Lemma \ref{THM:secondary_rate_adhoc}, and Corollary \ref{THM:clusters_of_preservation_regions}, respectively.

\begin{figure}
  \begin{center}
  \scalebox{0.9}{\includegraphics{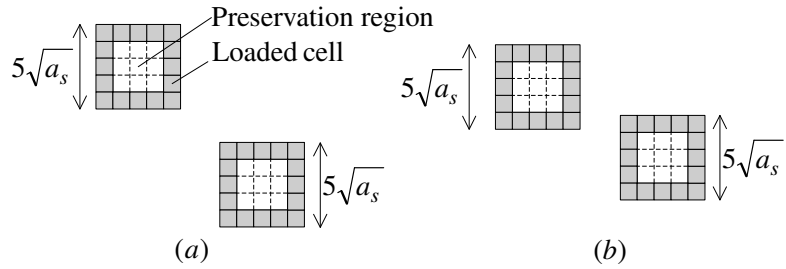}}
  \caption{An upper bound on the number of secondary S-D pairs whose extended HDPs pass through the loaded cells.}
  \label{FIG:high_loaded}
  \end{center}
\end{figure}

\begin{proof}
Note that by setting $\delta_P\in(0,\min\{\delta_{P,\operatorname{max}},1\})$, the secondary network satisfies (\ref{EQ:fraction_loss}) during its transmission.
Let us first consider $\beta>4/3$.
Let $m_h$ (similarly, $m_v$) denote the number of secondary
S-D pairs whose original or re-routed HDPs (VDPs) pass through loaded cells.
Suppose the following two cases where the projections of two preservation regions on the $y$-axis are at a distance greater than $2\sqrt{a_s}$ (Fig. \ref{FIG:high_loaded}. (a)) and less than $2\sqrt{a_s}$ (Fig. \ref{FIG:high_loaded}. (b)), respectively.
For the first case, all extended HDPs in the area of $1\times 10\sqrt{a_s}$ will pass through the loaded cells generated by two preservation regions.
But for the second case, the number of extended HDPs passing through the loaded cells is less than the previous case w.h.p. because the corresponding area is smaller than $1\times 10\sqrt{a_s}$.
Thus, assuming that projections of all preservation regions on the $y$-axis are at a distance of at least $2\sqrt{a_s}$ from each other gives an upper bound on $m_h$.
In this worst-case scenario, all sources located in the area of $1\times5(1+\epsilon)n\sqrt{a_s}$ generate extended HDPs w.h.p., which must pass through the loaded cells, where we use the fact that the number of preservation regions is upper bounded by $(1+\epsilon)n$ w.h.p..
By assuming that all nodes are sources, the resulting upper bound follows $\mbox{Poisson}\left(\lambda=5(1+\epsilon)n^2\sqrt{a_s}\right)$.
Similarly, an upper bound on $m_v$ follows $\mbox{Poisson}\left(\lambda=5(1+\epsilon)n^2\sqrt{a_s}\right)$.
If $\beta>4$, we obtain
\begin{eqnarray} \label{EQ:prob_zero}
\mathbb{P}\left(m_h=0\right)=\mathbb{P}\left(m_v=0\right)\!\!\!\!\!\!\!\!\!&&=\frac{e^{-5(1+\epsilon)n^2\sqrt{a_s}}\left(5(1+\epsilon)n^2\sqrt{a_s}\right)^k}{k!}\Big|_{k=0}\nonumber\\
&&=e^{-5(1+\epsilon)\sqrt{2\beta}n^{2-\frac{\beta}{2}}\sqrt{\log n}}\rightarrow 1, \mbox{ as }n\rightarrow\infty.
\end{eqnarray}
If $4/3<\beta\leq4$, from Lemma \ref{THM:upper_bound_poission}, we obtain
\begin{equation} \label{EQ:upper_m_ah}
\mathbb{P}\left(m_h\geq 10(1+\epsilon)n^2\sqrt{a_s}\right)\leq e^{-5(1+\epsilon)n^2\sqrt{a_s}}\left(\frac{e}{2}\right)^{10(1+\epsilon)n^2\sqrt{a_s}}.
\end{equation}
Then,
\begin{eqnarray} \label{EQ:upper_m_ah_av}
&&\mathbb{P}\left(m_h+m_v\geq 20(1+\epsilon)n^2\sqrt{a_s}\right)\nonumber\\
&&\leq\mathbb{P}\left((m_h\geq 10(1+\epsilon)n^2\sqrt{a_s})\cup(m_v\geq 10(1+\epsilon)n^2\sqrt{a_s})\right)\nonumber\\
&&\leq 2e^{-5(1+\epsilon)n^2\sqrt{a_s}}\left(\frac{e}{2}\right)^{10(1+\epsilon)n^2\sqrt{a_s}}\rightarrow0\mbox{ as }n\rightarrow\infty.
\end{eqnarray}
Hence, if $\beta>4/3$, we obtain w.h.p.
\begin{equation}
m_h+m_v\leq\epsilon_{s,2}(m)(1-\epsilon)\frac{m}{2},
\end{equation}
where $\epsilon_{s,2}(m)=40\sqrt{2}\frac{1+\epsilon}{1-\epsilon}\frac{\sqrt{\log m}}{m^{3/2-2/\beta}}$.
In conclusion, the fraction of S-D pairs whose
data paths pass through the loaded cells is upper bounded by
$\epsilon_{s,2}(m)$ w.h.p., which tends to zero as $m\to\infty$.
This indicates that almost all data paths will pass through regular cells rather than loaded cells.
If we treat the S-D pairs passing through the loaded cells and the S-D pairs not served as outages, $\epsilon_s(m)$ is obviously upper bounded w.h.p. by
\begin{equation}
\epsilon_s(m)\leq \epsilon_{s,1}(m)+\epsilon_{s,2}(m)=\Theta\left(\frac{\log m}{m^{1-1/\beta}}+\frac{\sqrt{\log m}}{m^{3/2-2/\beta}}\right),
\end{equation}
where we use the fact that the fraction of S-D pairs not served is upper bounded by $\epsilon_{s,1}(m)$ w.h.p. (Lemma \ref{THM:number_of_served_SD_adhoc}).
Then the achievable per-node throughput is determined by the rate of S-D pairs passing only the regular cells.
Since each secondary cell can sustain a constant rate of $K_s$ w.h.p. (Lemma \ref{THM:secondary_rate_adhoc}), from the result of Lemma \ref{THM:secondary_traffic_adhoc}, each served secondary S-D pair that passes only through regular cells can achieve a rate of at least $\frac{K_s}{4\sqrt{2m\log m}}$ w.h.p..
Therefore, $T_s(m)$ is lower bounded by $\frac{K_s}{4\sqrt{2}}\frac{1}{\sqrt{m\log m}}$ w.h.p..

Let us now consider the case when $1<\beta\leq4/3$.
Unlike the previous case, most served S-D pairs in this case pass
through loaded cells, which will become bottlenecks.
By assuming that all served S-D pairs pass through loaded cells, we obtain a lower bound on $T_s(m)$ with $\epsilon_s(m)\leq \epsilon_{s,1}(m)=\Theta\left(\frac{\log m}{m^{1-1/\beta}}\right)$, which is an upper bound on the fraction of unserved S-D pairs.
Therefore, based on Lemmas  \ref{THM:secondary_rate_adhoc} and \ref{THM:secondary_traffic_adhoc}, $T_s(m)$
is lower bounded by $\frac{K_s}{4(6N_c+1)\sqrt{2m\log m}}$ w.h.p..

Since there are at least $(1-\epsilon)(1-\epsilon_s(m))\frac{m}{2}$ non-outage S-D pairs, $S_s(m)$ is lower bounded by $(1-\epsilon)(1-\epsilon_s(m))\frac{m}{2}T_s(m)$ w.h.p., which completes the proof.
\end{proof}

\begin{figure}[t!]
  \begin{center}
  \scalebox{0.6}{\includegraphics{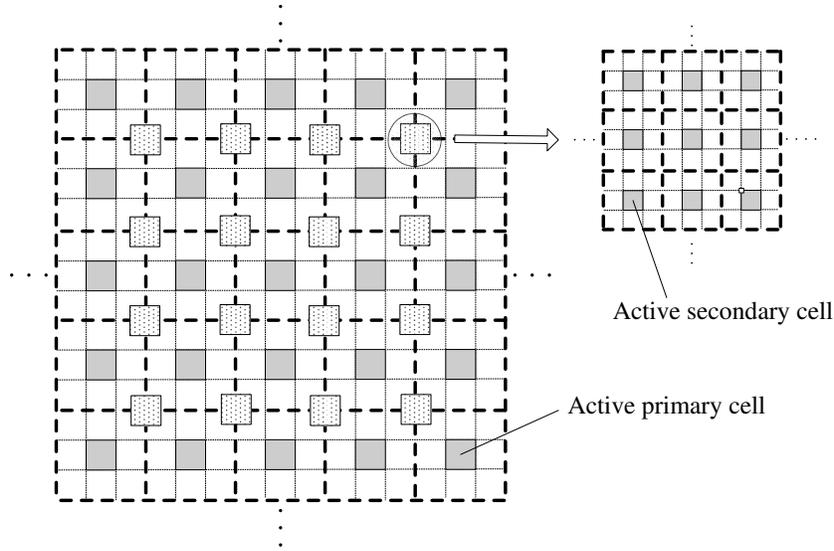}}
  \caption{Alternative secondary protocol with different information about the primary network: the secondary network operates based on $81$-TDMA.}
  \label{FIG:alternative_adhoc}
  \end{center}
\end{figure}

Notice that if the secondary network knows when the primary nodes are activated in addition to their location, then $81$-TDMA between the secondary cells in Fig. \ref{FIG:alternative_adhoc} can achieve the same scaling laws of Theorem \ref{THM:secondary_throughput_adhoc}.
Specifically, each group of the secondary cells can be activated based on the $9$-TDMA (dotted region) and within each group secondary cells operate $9$-TDMA.


\subsection{Indirect Multihop Routing for the Primary Network} \label{subsec:indirect_routing}
\subsubsection{Indirect multihop routing protocol}
The indirect multihop routing in \cite{Franceschetti:07} can also be adopted as a primary protocol, which provides the sum throughput of $\Theta(\sqrt{n})$.
The key observation is that the construction of multihop data paths with a hop distance of $\Theta(1/\sqrt{n})$ is possible, which consists of the ``highway'' for multihop transmission.
During Phase 1, each source directly transmits its packet to the closest node on the highway and, during Phase 2, the packet is delivered to the node on the highway closest to the destination by multihop transmissions using the nodes on the highway.
Finally, during Phase 3, the destination directly receives the packet from the closet node on the highway.

\subsubsection{Throughput scaling laws} Let us assume that the transmit power of each primary Tx scales according to the hop distance, that is each primary Rx will receive the intended signal with a constant power.
Since the hop distance for Phase 1 (or 3) is given by $\Theta(\log n/\sqrt{n})$, which is greater than $\Theta(\sqrt{\log n/n})$ achieved by the direct routing, the transmit power of Phase 1 (or 3) is greater than that of the direct routing.
The transmit power of Phase 2, on the other hand, is smaller than that of the direct routing because the hop distance is given by $\Theta(1/\sqrt{n})$.
Therefore, we can apply the previous secondary routing protocol during Phase 2 of the primary indirect routing, which will cause less interference to the secondary network.
Based on the analysis used for the direct routing, we derive the same results of Theorems \ref{THM:primary_throughput_adhoc} and \ref{THM:secondary_throughput_adhoc} except now we have $T_{\operatorname{alone}}(n)=\Theta(1/\sqrt{n})$ and $S_{\operatorname{alone}}(n)=\Theta(\sqrt{n})$.

\section{Infrastructure-Supported Primary Network}
\label{sec:infra}

In this section, we consider a different primary network which includes additional regularly-spaced BSs.
Here the primary nodes are again randomly distributed over a
given area according to a p.p.p. of density $n$. In
addition, the communication between the primary nodes is aided by the
presence of $l$ BSs, which may
communicate at no cost in terms of scaling. In this
infrastructure-supported primary model, the secondary network
continues to operate in an ad hoc fashion with nodes distributed
according to a p.p.p. of density $m=n^{\beta}$. Again
we consider $\beta>1$ only.

We first outline the main results before describing the network protocols and analyzing the throughput and its asymptotic behavior for both the primary and secondary networks.

\subsection{Main Results}


{\em
Suppose the infrastructure-supported primary model with $\gamma > 1/2$.
For any $\delta_{\operatorname{loss}}>0$, the primary network can achieve the following per-node and sum throughputs w.h.p.:
\begin{equation}
T_p(l)=(1-\delta_{\operatorname{loss}})T_{\operatorname{alone}}(l), \hspace{2cm} S_p(l)=(1-\delta_{\operatorname{loss}})S_{\operatorname{alone}}(l),
\end{equation}
where $T_{\operatorname{alone}}(l)=\Theta(l^{1-1/\gamma})$ and $S_{\operatorname{alone}}(l)=\Theta\left(l\right)$.
The following per-node and sum throughputs are $\epsilon_s(m)$-achievable w.h.p. for the secondary network:
\begin{equation}
T_s(m)=\Theta\left(\sqrt{\frac{1}{m\log m}}\right), \hspace{2cm} S_s(m)=\Theta\left(\sqrt{\frac{m}{\log m}}\right)
\end{equation}
where  $\epsilon_s(m)=\operatorname{O}(1/\log m)$, which converges to zero as $m\to\infty$.
}

Compared to the throughput scalings of the ad hoc primary model, the addition of BSs helps increase the scaling law
of the primary network if $\gamma > 1/2$, otherwise the scaling law
stays unaffected \cite{Liu:03}. We show here that the presence of a secondary
network does not change the scaling law of this primary
network for $\gamma > 1/2$ (For $\gamma \leq 1/2$, the results of the
previous ad hoc primary model apply). The secondary
network can again achieve, with an arbitrarily small fraction of outage, the same scaling law under the direct multihop routing
protocol as when the primary network is absent.

\subsection{Network Protocols}
We assume the primary network uses a classical BS-based data transmission, in which sources deliver data to BSs during the uplink phase and BSs deliver received data to destinations during the downlink phase.
The challenge is again to prove that the secondary nodes can transmit in such a way that the primary scaling law should satisfy $T_p(l)\geq (1-\delta_{\operatorname{loss}})T_{\operatorname{alone}}(l)$ w.h.p..

\subsubsection{Primary network protocol}
We consider the primary protocol in which a
source node transmits a packet to its closest BS and the destination
node receives the packet from its closest BS, similar to those in \cite{Liu:03} and \cite{Zemlianov:05}:
\begin{itemize}
\item Divide the unit area into square primary cells of  area $a'_p=\frac{1}{l}$, where each primary cell has one BS at its center.
\item During the uplink phase, each source node transmits a packet to the closest BS.
\item The BS that receives a packet from a source delivers it to the BS closest to the corresponding destination using BS-to-BS links.
\item During the downlink phase, each destination node receives its packet from the closest BS.
\item A simple round-robin scheme is used for all downlink transmissions and all uplink transmissions in the same primary cell.
\item At each transmission, a Tx node transmits with power $Pd^{\alpha}$, where $d$ denotes the distance between the Tx and its Rx.
\end{itemize}

Under the given primary protocol, the sum throughput of $S_{\operatorname{alone}}(l)=\Theta\left(l\right)$ is achievable, which coincides with the result of \cite{Liu:03}.
Note that if $\gamma>1/2$, $S_{\operatorname{alone}}(l) = \Theta(l)>\Theta\left(\sqrt{n}\right)$.
That is, when $\gamma>1/2$, using BSs helps improve the
throughput scaling of the primary network.
As was pointed out in \cite{Liu:03}, to improve throughput scaling,
the number of BSs should be high enough.
Therefore, this primary protocol for the infrastructure-supported
model is suitable for $\gamma>/1/2$, while the result of the
ad hoc primary model can be applied for $0<\gamma\leq1/2$.

\subsubsection{Secondary network protocol}
Let us consider the secondary protocol when the primary network is in the downlink phase.
Since the secondary cell size is smaller than the primary cell size, the amount of interference from the secondary network to the primary network may be reduced by setting a preservation region around each primary receiving node.
However, the repeated transmissions of the same secondary packet does not guarantee a non-trivial rate for secondary transmissions since all BSs are always active in the worst case for the infrastructure-supported case.
Similar to the concept of preservation regions, in order to reduce the interference to the secondary nodes, in a certain region around each BS (which are primary Txs) we insist that no secondary nodes transmit or receive in that region.
However, due to the relatively high transmit power of primary
transmissions, these regions need a larger area than the previously defined
preservation region.
Define an \emph{avoidance region} as a square
containing $\delta_a\frac{a'_p}{a'_s}$ secondary cells with a BS at the center, where $a'_s$ is the size of the secondary cell that is the same as $a_s$.
We also set the preservation regions around each BS consisted of $\frac{\delta_a}{\log n}\frac{a'_p}{a'_s}$ secondary cells and around each primary node consisted of $9$ secondary cells.
We obtain a secondary protocol by replacing the three repeated transmissions of the previous secondary protocol by:
\begin{itemize}
\item If a horizontal or vertical data path of each secondary S-D pair is blocked by an avoidance region, this data path is shifted horizontally (or vertically) to the non-blocked region.
\item Divide the entire time into two phases, where $\delta_t\in(0,1)$ denotes the time fraction for Phase $1$.
During Phase $1$, Txs in the avoidance regions perform multihop transmissions using $\delta_t$ time fraction.
During Phase $2$, Txs outside the avoidance regions perform multihop transmissions using $1-\delta_t$ time fraction.
\end{itemize}

\begin{figure}
  \begin{center}
  \scalebox{0.85}{\includegraphics{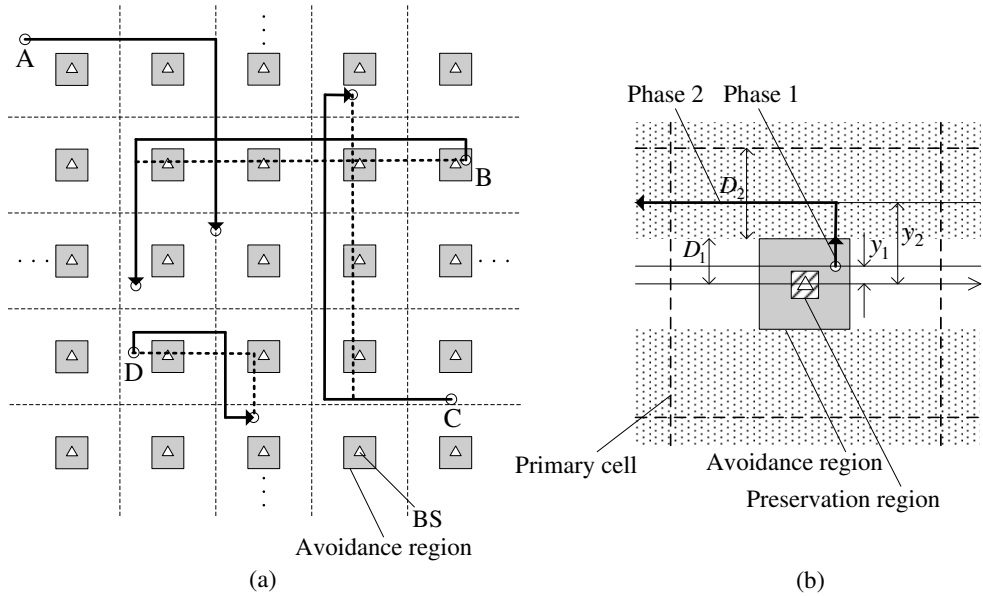}}
  \caption{Secondary data paths for the infrastructure-supported primary model: a horizontal (or vertical) data path is horizontally (or vertically) shifted if it is blocked by an avoidance region. The dotted regions denoted by $\mathcal{R}_h$ are the regions in which data paths are free from avoidance regions.}
  \label{FIG:secondary_routing_infra}
  \end{center}
\end{figure}

Fig. \ref{FIG:secondary_routing_infra}. (a) illustrates
examples of shifted secondary data paths due to the avoidance regions (for simplicity,
preservation regions are not shown in this figure): $A$ illustrates the case
where the  HDP and VDP are not blocked, $B$ the case where only the HDP
is blocked, $C$ the case where only  the  VDP is blocked, and
$D$ the case where both the HDP and VDP are blocked.
Fig. \ref{FIG:secondary_routing_infra}. (b) illustrates the shifted HDP of the case $B$.
Since the source is in the avoidance region (but not in the preservation region), the multihop from the source to the first receiving node outside the avoidance region will be conducted during Phase $1$ and the rest multihop to the destination will be conducted during Phase $2$.

\bigskip
{\bf Avoidance region re-routing:}

Since the area of each avoidance region is much larger than that of each preservation region, secondary cells adjacent to the avoidance regions should handle much more traffic than regular cells if we were to  re-route blocked data paths using \emph{only} these cells.
In order to more evenly  distribute the re-routed traffic, we shift an entire data path to the non-blocking region based on given mapping rule for the case when it is blocked by an avoidance region.
Let us consider the details of finding a shifted secondary data path when it is blocked by an avoidance region.
Define $\mathcal{R}_h$ as the region in which extended HDPs are not blocked by the avoidance regions. This region is guaranteed to exist because of the regular BS placement, which is
shown by the dotted regions in Fig. \ref{FIG:secondary_routing_infra}. (b).
Let us focus on the case $B$, where the blocked HDP in $\mathcal{R}^c_h$ is shifted to the new HDP in $\mathcal{R}_h$.
Let $y_1$ and $y_2$ denote the $y$-axis of the blocked HDP and of its shifted HDP, respectively.
Without loss of generality, it is assumed that $y_1$ is in $\left[0,D_1\right]$, where $D_1=\frac{1}{2}\sqrt{\delta_a a'_p}$.
Then $y_2$ is given by
\begin{equation}
y_2=\frac{D_2}{D_1}y_1+D_1,
\end{equation}
where $D_2=\frac{1}{2}\sqrt{a'_p}-\frac{1}{2}\sqrt{\delta_aa'_p}$.
Note that $D_1$ is half of the side length of an avoidance region, while $D_2$ is half of the length of the strips which are free of avoidance regions.
Similarly, let  $\mathcal{R}_v$ denote the region in which none of VDPs are blocked.
We can shift a blocked VDP in $\mathcal{R}^c_v$ to $\mathcal{R}_v$ using the analogous mapping to the horizontal case.
If a HDP is shifted, it requires  a series of short vertical hops to reach the shifted HDP, where we denote these vertical hops as a short VDP.
It also requires short horizontal hops to reach a destination after the VDP if that VDP is shifted, where we denote these horizontal hops as a short HDP.

\bigskip

Let us consider the secondary protocol when the primary network is in the uplink phase.
We can also define an avoidance region at each Tx (primary node) of the primary network.
Due to the irregular placement of primary nodes, however, it is hard to construct a re-routing protocol when each data path is blocked by an avoidance region.
More importantly, we cannot set the area of each avoidance region as large as in the downlink case since the density of primary nodes is higher than that of BSs, leading to a  smaller throughput than the downlink case.
Note that if we operate the secondary network during the uplink and downlink phases separately, then throughput scalings of the secondary network follow the maximum of the uplink and downlink throughputs.
Therefore, overall throughput scalings follow those of the downlink phase.

\subsection{Throughput Analysis and its Asymptotic Behavior}
In this subsection, we analyze the per-node and sum
throughputs of each network under given protocols and derive the
corresponding scaling laws.

\subsubsection{Primary network throughputs}

Let us consider the per-node throughput of the primary network in the presence of the secondary network.
We first show that all primary cells may sustain a constant, non-trivial rate in Lemma \ref{THM:primary_rate_infra}.
We then determine the number of uplink and downlink transmissions each of these cells must support in Lemma \ref{THM:primary_traffic_infra}.
Using these results, we obtain the primary per-node and sum throughputs in Theorem \ref{THM:primary_throughput_infra}.

%
Let $R'_p(l)$ and $R'_{\operatorname{alone}}(l)$ denote the achievable aggregate rate of each primary cell in the presence and in the absence of the secondary network, respectively.
We define
\begin{equation}
I'\triangleq P2^{\alpha/2+3}\sum_{t=1}^{\infty}t(2t-1)^{-\alpha}
\label{EQ:def_I'}
\end{equation}
having a finite value for $\alpha>2$, which will be used to derive an upper bound on the interference power of the infrastructure-supported primary network.
Then the following lemma holds.

\begin{lemma} \label{THM:primary_rate_infra}
Suppose the infrastructure-supported model.
If $\delta_P\in(0,\min\{\delta'_{P,\operatorname{max}},1\})$, then
\begin{equation}
\lim_{l\to\infty}\frac{R'_p(l)}{R'_{\operatorname{alone}}(l)}\geq 1-\delta_{\operatorname{loss}},
\label{EQ:R_p_R_alone}
\end{equation}
where $\delta'_{P,\max}=\big(\frac{1}{(1+\frac{P}{N_0})^{1-\delta_{\operatorname{loss}}}-1}-\frac{N_0}{P}\big)\frac{P}{I}$ and $I'$ is given by (\ref{EQ:def_I'}).
Moreover, $R'_{\operatorname{alone}}(l)$ is lower bounded by $K'_p$, where $K'_p=\log\left(1+\frac{P}{N_0+I'}\right)$ is a constant independent of $l$.
\end{lemma}
\begin{proof}
The proof is in the Appendix.
\end{proof}


\begin{lemma} \label{THM:primary_traffic_infra}
Under the infrastructure-supported model, each primary cell needs to carry at most $2n^{1-\gamma}$ downlink
and $2n^{1-\gamma}$ uplink transmissions w.h.p..
\end{lemma}
\begin{proof}
The proof is in the Appendix.
\end{proof}

\begin{theorem} \label{THM:primary_throughput_infra}
Suppose the infrastructure-supported model.
For any $\delta_{\operatorname{loss}}\in(0,1)$, by setting $\delta_P\in(0,\min\{\delta'_{P,\operatorname{max}},1\})$, the primary network can achieve $T_p(l)=(1-\delta_{\operatorname{loss}})T_{\operatorname{alone}}(l)$
and $S_p(l)=(1-\delta_{\operatorname{loss}})S_{\operatorname{alone}}(l)$ w.h.p., where
\begin{equation}
T_{\operatorname{alone}}(l)=\frac{K'_p}{4}l^{1-\frac{1}{\gamma}}
\end{equation}
and
\begin{equation}
S_{\operatorname{alone}}(l)=\frac{K'_p(1-\epsilon)}{8}l.
\end{equation}
The definitions of $\delta'_{P,\operatorname{max}}$ and $K'_p$ are given in Lemma \ref{THM:primary_rate_infra}.
\end{theorem}
\begin{proof}
First consider the stand-alone throughput of the primary network.
Let $T_{\operatorname{alone},d}(l)$ and $T_{\operatorname{alone},u}(l)$ denote the per-node throughput during downlink and uplink, respectively.
Then $T_{\operatorname{alone}}(l)=\frac{1}{2}\min\left\{T_{\operatorname{alone},d}(l), T_{\operatorname{alone},u}(l)\right\}$, where $\frac{1}{2}$ arises from the fact that a source delivers a packet to its destination using one downlink and one uplink transmission.
Since each primary cell can sustain a constant rate of $K'_p$ (Lemma
\ref{THM:primary_rate_infra}), $T_{\operatorname{alone},d}(l)$ is upper bounded by $K'_p$
divided by the maximum number of downlink transmissions in each primary cell.
This number of downlink transmissions is upper bounded by
$2n^{1-\gamma}$ w.h.p. (Lemma \ref{THM:primary_traffic_infra}).
Therefore, $T_{\operatorname{alone},d}(l)$ is lower bounded by $\frac{K'_p}{2n^{1-\gamma}}$ w.h.p..
Since the same lower bound can be obtained for the case of $T_{\operatorname{alone},u}(l)$, $T_{\operatorname{alone}}(l)$ is lower bounded by $\frac{K'_p}{4n^{1-\gamma}}$ w.h.p..
From the fact that there are at least $(1-\epsilon)\frac{n}{2}$ primary S-D
pairs (Lemma \ref{THM:node_number}),
$S_{\operatorname{alone}}(n)$ is lower bounded by $(1-\epsilon)\frac{n}{2}T_p(n)$ w.h.p..
The remaining proof about $T_p(l)=(1-\delta_{\operatorname{loss}})T_{\operatorname{alone}}(l)$
and $S_p(l)=(1-\delta_{\operatorname{loss}})S_{\operatorname{alone}}(l)$ w.h.p. is the same as Theorem \ref{THM:primary_throughput_adhoc}, which completes the proof.
\end{proof}

\subsubsection{Secondary network throughputs}
Let us now consider the throughput scalings of the secondary network in the presence of the primary network.
We first show that the fraction of the unserved S-D pairs due to the preservation regions will be negligible w.h.p. in Lemma \ref{THM:number_of_served_SD_infra}.
Unlike the ad hoc primary model, the overall multihop transmission of each S-D pair is divided into Phases 1 and 2 depending on each Tx's location.
Hence the per-node throughput scales as the minimum of the rate scalings related to Phases 1 and 2, respectively.
We will show that although the aggregate rate of each secondary cell in the avoidance regions decreases as $\Theta(\log m)^{-\alpha/2}$ (Lemma \ref{THM:secondary_rate_infra}), the number of data paths delivered by this cell is much less than that of each secondary cell outside the avoidance regions (Lemmas \ref{THM:secondary_traffic_infra} and \ref{THM:secondary_traffic_infra2}).
Thus the cells in the avoidance regions provide higher rate per each hop transmission than the cells outside the avoidance regions w.h.p. and, as a result, $T_s(m)$ and $S_s(m)$ are determined by the transmissions outside the avoidance regions, which is Phase $2$.

\begin{lemma} \label{THM:number_of_served_SD_infra}
Under the infrastructure-supported primary model, the fraction of unserved secondary S-D pairs is upper
bounded by $\epsilon'_{s,1}(m)= \Theta(1/\log m)$ w.h.p., which converges to zero as $m\to \infty$.
\end{lemma}
\begin{proof}
The proof is in the Appendix.
\end{proof}

\begin{lemma} \label{THM:secondary_rate_infra}
Under the infrastructure-supported primary model, each secondary cell in the avoidance regions and each secondary cell outside the avoidance regions can sustain a rate of $K'_{s,1}(m)$ and $K'_{s,2}$ respectively,
where $K'_{s,1}(m)=\frac{\delta_t}{18}\log\left(1+\frac{\delta_P P}{N_0+I'+\delta_P I+P\left(2\log m/(\beta\delta_a)\right)^{\alpha/2}}\right)$, which tends to zero as $m\to\infty$,
and $K'_{s,2}=\frac{1-\delta_t}{18}\log\left(1+\frac{\delta_PP}{N_0+I'+\delta_PI+P\left(2/\delta_a\right)^{\alpha/2}}\right)$ is a constant independent of $m$.
The definitions of $I$ and $I'$ are given by (\ref{EQ:def_I}) and (\ref{EQ:def_I'}), respectively.
\end{lemma}
\begin{proof}
The proof is in the Appendix.
\end{proof}


As in the ad hoc primary model, we define the secondary cells which border the preservation regions as the \emph{loaded} cells and the other cells as \emph{regular} cells.
Then, the following lemmas hold.

\begin{lemma} \label{THM:secondary_traffic_infra}
Suppose the infrastructure-supported primary model.
Each regular secondary cell and each loaded secondary cell outside the avoidance regions need to carry at most
$4(1-\sqrt{\delta_a})^{-1}\sqrt{2m\log m}$ and $4\left(6N_c+1\right)(1-\sqrt{\delta_a})^{-1}\sqrt{2m\log m}$ data paths w.h.p., respectively, where $N_c$ is given in Corollary \ref{THM:clusters_of_preservation_regions}.
\end{lemma}
\begin{proof}
The proof is in the Appendix.
\end{proof}

\begin{lemma} \label{THM:secondary_traffic_infra2}
Suppose the infrastructure-supported primary model.
Each regular secondary cell and each loaded secondary cell in the avoidance regions need to carry at most
$2\sqrt{2\delta_am^{1-\gamma/\beta}\log m}$ and $2\left(6N_c+1\right)\sqrt{2\delta_am^{1-\gamma/\beta}\log m}$ data paths w.h.p., respectively, where $N_c$ is given in Corollary \ref{THM:clusters_of_preservation_regions}.
\end{lemma}
\begin{proof}
The proof is in the Appendix.
\end{proof}

We can now use the previous corollaries and lemmas to obtain the
per-node and sum throughputs of the secondary network in the following
theorem.

\begin{theorem} \label{THM:secondary_throughput_infra}
Suppose the infrastructure-supported primary model.
For any $\delta_{\operatorname{loss}}\in(0,1)$, by setting $\delta_P$ within $(0,\min\{\delta'_{P,\operatorname{max}},1\})$,
the following per-node and sum throughputs are $\epsilon_s(m)$-achievable for the secondary network w.h.p.:
\begin{equation}
T_s(m)=\begin{cases} \frac{K'_{s,2}}{4\sqrt{2}(1-\sqrt{\delta_a})^{-1}}\frac{1}{\sqrt{m\log m}}&\text{if $\beta>\frac{4}{3}$}\\
\frac{K'_{s,2}}{4\sqrt{2}(1-\sqrt{\delta_a})^{-1}(6N_c+1)}\frac{1}{\sqrt{m\log m}}&\text{if $1<\beta\leq\frac{4}{3}$}\end{cases}
\end{equation}
and
\begin{equation}
S_s(m)=\begin{cases} \frac{K'_{s,2}(1-\epsilon)(1-\epsilon_s(m))}{8\sqrt{2}(1-\sqrt{\delta_a})^{-1}}\sqrt{\frac{m}{\log m}}&\text{if $\beta>\frac{4}{3}$}\\
\frac{K'_{s,2}(1-\epsilon)(1-\epsilon_s(m))}{8\sqrt{2}(1-\sqrt{\delta_a})^{-1}(6N_c+1)}\sqrt{\frac{m}{\log m}}&\text{if $1<\beta\leq\frac{4}{3}$},
\end{cases}
\end{equation}
where $\epsilon_s(m)=\operatorname{O}(\frac{1}{\log m})$, which converges to zero as $m\to\infty$.
The definitions of $\delta'_{P,\operatorname{max}}$, $K'_{s,2}$, and $N_c$ are given in Lemma \ref{THM:primary_rate_infra}, Lemma \ref{THM:secondary_rate_infra}, and Corollary \ref{THM:clusters_of_preservation_regions}, respectively.
\end{theorem}

\begin{proof}
Note that by setting $\delta_P\in(0,\min\{\delta'_{P,\operatorname{max}},1\})$, the secondary network satisfies (\ref{EQ:fraction_loss}) during its transmission.
Let us first consider $\beta>4/3$.
Let $m'_h$ (similarly, $m''_h$) denote the number of secondary
S-D pairs whose original, including shifted one, or re-routed HDPs are in $\mathcal{R}_h$
($\mathcal{R}^c_h$) and pass through loaded
cells. Similarly, we can define $m'_v$ and $m''_v$ for extended VDPs.

To obtain an upper bound on $m'_h$, we consider extended HDPs, which is the same as Lemma \ref{THM:secondary_traffic_infra}, and study the geometric
scenario that requires re-routing the largest number of data paths to
the loaded cells. This worst-case scenario is obtained when the
projections of all preservation regions on the $y$-axis are separated
at a distance of at least $2\sqrt{a'_s}$ and all preservation regions
are in the avoidance-region free zone $\mathcal{R}_h$. Thus, all nodes
located in the area of $1\times5c(1+\epsilon)n\sqrt{a'_s}$ pass
through loaded cells, where $c=(1-\sqrt{\delta_a})^{-1}$ arises from the
shifted HDPs along with the original HDPs.
Therefore, an upper bound on $m'_h$ follows $\mbox{Poisson}\left(5c(1+\epsilon)n^2\sqrt{a'_s}\right)$.
Similarly, an upper bound on $m''_h$ follows $\mbox{Poisson}\left(5(1+\epsilon)n^2\sqrt{a'_s}\right)$, where we assume that all preservation regions are in $\mathcal{R}^c_h$ for this case.
The vertical worst-case scenario may be similarly derived.
Using the same analysis from (\ref{EQ:prob_zero}) to (\ref{EQ:upper_m_ah_av}), we obtain w.h.p.
\begin{equation}
m'_{a,h}+m'_{a,v}+m''_{a,h}+m''_{a,v}\leq\epsilon'_{s,2}(m)(1-\epsilon)\frac{m}{2},
\end{equation}
where $\epsilon'_{s,2}(m)=40\sqrt{2}(1+c)\frac{1+\epsilon}{1-\epsilon}\frac{\sqrt{\log m}}{m^{3/2-2/\beta}}$.
If we treat the S-D pairs passing through the loaded cell and the S-D pairs not served as outage,
\begin{equation}
\epsilon_s(m)\leq \epsilon'_{s,1}(m)+\epsilon'_{s,2}(m)=\Theta\left(1/\log m\right)
\end{equation}
w.h.p., where we use the result of Lemma \ref{THM:number_of_served_SD_infra}.
Then the achievable per-node throughput is determined by the rate of S-D pairs passing through only the regular cells.
Let us consider the regular cells in the avoidance regions, which perform transmissions during Phase $1$.
For this case, since each cell sustains a rate of $K'_{s,1}(m)$ w.h.p. (Lemma \ref{THM:secondary_rate_infra}), and based on Lemma \ref{THM:secondary_traffic_infra2},
the rate per each hop transmission provided by these cells is lower bounded by
\begin{equation}
\frac{K'_{s,1}(m)}{2\sqrt{2\delta_am^{1-\gamma/\beta}\log m}}=\Theta\left(\frac{(\sqrt{\log m})^{-\alpha}}{\sqrt{m^{1-\gamma/\beta}\log m}}\right)
\end{equation}
w.h.p..
If we consider the regular cells outside the avoidance regions, from Lemmas \ref{THM:secondary_rate_infra} and \ref{THM:secondary_traffic_infra},
the rate per each hop transmission is lower bounded by
\begin{equation}
\frac{K'_{s,2}}{4(1-\sqrt{\delta_a})^{-1}\sqrt{2m\log m}}=\Theta\left(\frac{1}{\sqrt{m\log m}}\right)
\end{equation}
w.h.p..
Since, for sufficiently large $m$, the rate provided by the cells in the avoidance regions is greater than that provided by the cells outside the avoidance regions, $T_s(m)$ is lower bounded by $\frac{K'_{s,2}}{4\sqrt{2}(1-\sqrt{\delta_a})^{-1}}\frac{1}{\sqrt{m\log m}}$ w.h.p. if $\beta>4/3$.

Let us now consider $1<\beta\leq4/3$.
Again, we obtain a lower bound on $T_s(m)$ by considering the most heavily loaded scenario in which all served S-D pairs pass through loaded cells.
Then $\epsilon_s(m)\leq \epsilon'_{s,1}(m)=\Theta(1/\log m)$.
Similarly, we can derive the rate per each hop transmission related to Phases $1$ and $2$ from the results in Lemmas \ref{THM:secondary_rate_infra} to \ref{THM:secondary_traffic_infra2}.
As a result, $T_s(m)$ is lower bounded by $\frac{K'_{s,2}}{4\sqrt{2}(1-\sqrt{\delta_a})^{-1}(6N_c+1)}\frac{1}{\sqrt{m\log m}}$ w.h.p. if $1<\beta\leq4/3$.

Finally $S_s(m)$ is lower bounded by $(1-\epsilon)(1-\epsilon_s(m))\frac{m}{2}T_s(m)$ w.h.p., which completes the proof.
\end{proof}

\section{Conclusion}
\label{sec:conclusion}
In this paper, we studied two co-existing ad hoc networks with different priorities (a primary and a secondary network)  and analyzed their simultaneous  throughput scalings.
It was shown that each network can achieve the same throughput scaling as when the other network is absent.
Although we allow outage for the secondary S-D pairs, the fraction of pairs in outage converges to zero  as node densities increase.
Furthermore, these scalings may be achieved by adjusting the secondary protocol while keeping that of the primary network unchanged. In essence, the primary network is unaware of the presence of the secondary network.  To achieve this result, the secondary nodes need knowledge of the locations of the primary nodes, and the secondary nodes need to be denser than the primary.
For $\beta\leq1$ (primary is denser than the secondary network), on the other hand, it seems to be more challenging to achieve similar throughput scaling results while keeping the primary unchanged, as there are many primary nodes around each secondary node.
As mentioned before, if we allow  the primary protocol to adapt to the presence of the secondary network, we can achieve throughput scalings of two homogenous networks by employing TDMA between two networks.
Our result may be extended to more than two networks, provided each layered network obeys the same three main assumptions as in the two network case.


\appendix


Before proving our lemmas,  we recall
the following useful lemma from \cite{Franceschetti:07}.

\begin{lemma}[Franceschetti, Dousse, Tse, and Thiran] \label{THM:upper_bound_poission}
Let $X$ be a Poisson random variable with parameter $\lambda$. Then
\begin{equation}
\mathbb{P}(X\geq x)\leq \frac{e^{-\lambda}(e \lambda)^x}{x^x},{~~}\text{for}{~~}x>\lambda.
\end{equation}
\end{lemma}
\begin{proof}
We refer readers to the paper \cite{Franceschetti:07}.
\end{proof}

\bigskip

\noindent{\it Proof of Lemma \ref{THM:node_number}}

Let $X_1$ denote the number of primary nodes in a unit area.
For part (a), we wish to show that $\mathbb{P}(|X_1-n|\geq \epsilon \, n) \rightarrow 0$ as $n\rightarrow \infty$. Noting that $X_1$ is a Poisson random variable with mean $n$ and standard deviation $\sqrt{n}$,
we use Chebyshev's inequality to see that
\begin{align*}
\mathbb{P}\left( |X_1-n| \geq (\epsilon \sqrt{n})\sqrt{n}\right) &\leq \frac{1}{(\epsilon\sqrt{n})^2}.
\end{align*}
Clearly, as $n$ tends to infinity we can make this quantity arbitrarily small.

For part (b), let $X_2$ denote the number of primary nodes in a primary cell.
Then $\mathbb{P}(X_2=0)$ is given by
\begin{equation}
\mathbb{P}(X_2=0)=\frac{e^{-2\log n}(2\log n)^k}{k!}\Big|_{k=0}=\frac{1}{n^2}.
\end{equation}
Therefore, the probability that there is at least one cell having no node is upper bounded by $n\mathbb{P}(X_2=0)$, where the union bound and the fact that there are at most $n$ primary cells are used.
Since $\frac{1}{n}\to0$ as $n\to\infty$, (b) holds w.h.p., which completes the proof.

\bigskip

\noindent{\it Proof of Lemma \ref{THM:primary_rate_adhoc1}}

\begin{figure}
  \begin{center}
  \scalebox{0.65}{\includegraphics{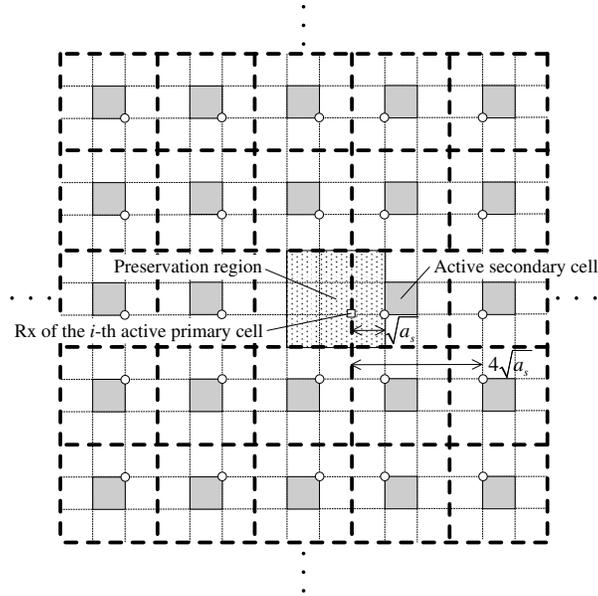}}
  \caption{The amount of interference from the secondary interferers to the Rx of the $i$-th primary pair for the ad hoc primary model, where the shaded cells indicate the active secondary cells based on the $9$-TDMA.}
  \label{FIG:i_p_adhoc}
  \end{center}
\end{figure}

Suppose that at a given moment, there are $N_p(n)$ active primary cells and $N_s(n)$ active secondary cells, including the $i$-th active primary cell.
Then, the rate of the $i$-th active primary cell is given by
\begin{equation}
R_p^i(n)=\frac{1}{9}\log\left(1+\frac{P_p^i
g\left(\|X_{p,\operatorname{tx}}^i-X_{p,\operatorname{rx}}^i\|\right)}{N_0+I_p^i(n)+I_{sp}^i(n)}\right),
\label{EQ:R_p_i}
\end{equation}
where $\frac{1}{9}$ indicates the loss in rate due to the $9$-TDMA transmission of primary cells.
The rate of the $i$-th active primary cell in the absence of the secondary network is given by $R_{\operatorname{alone}}^i(n)=R_p^i(n)$ by setting ${I_{sp}^i(n)=0}$.
Fig. \ref{FIG:i_p_adhoc} illustrates the worst case interference from the secondary interferers to the Rx of the $i$-th active primary cell, where the dotted region denotes the preservation region around the primary Rx and the shaded cells denote the active secondary cells based on the $9$-TDMA.
Because of the preservation region, the minimum distance of $\sqrt{a_s}$ can be guaranteed from all secondary transmitting interferers to the primary Rx.
Thus, there exist $8$ secondary interferers at a distance of at least $\sqrt{a_s}$, and $16$ secondary interferers at a distance of at least $4\sqrt{a_s}$, and so on.
Then, $I_{sp}^i(n)$ is upper bounded by
\begin{equation}
I_{sp}^i(n)=\sum_{k=1}^{N_s(n)}P_s^kg\left(\|X_{s,\operatorname{tx}}^k-X_{p,\operatorname{rx}}^i\|\right)<\delta_PP(\sqrt{2a_s})^{\alpha}\sum_{t=1}^{\infty}8t\left((3t-2)\sqrt{a_s}\right)^{-\alpha}=\delta_PI,
\label{EQ:upper_i_sp}
\end{equation}
where we use the fact that $P_s^k\leq \delta_P P(\sqrt{2a_s})^{\alpha}$.
Then
\begin{equation}
\lim_{n\to\infty}\frac{R^i_p(n)}{R^i_{\operatorname{alone}}(n)}\geq \lim_{n\to\infty}\frac{\log\left(1+\frac{P}{N_0+I^i_p(n)+\delta_PI}\right)}{\log\left(1+\frac{P}{N_0+I^i_p(n)}\right)}\geq \frac{\log\left(1+\frac{P}{N_0+\delta_PI}\right)}{\log\left(1+\frac{P}{N_0}\right)}.
\label{EQ:fraction_lower}
\end{equation}
Notice that $\delta_{P,\max}$ is the value of $\delta_P$ such that the right-hand side of (\ref{EQ:fraction_lower}) is equal to $1-\delta_{\operatorname{loss}}$.
Thus, if we set $\delta_P\in(0,\min\{\delta_{P,\operatorname{max}},1\})$, then $\lim_{n\to\infty}\frac{R^i_p(n)}{R^i_{\operatorname{alone}}(n)}\geq1-\delta_{\operatorname{loss}}$.
Because the above inequality holds for any $i$, we obtain $\lim_{n\to\infty}\frac{R_p(n)}{R_{\operatorname{alone}}(n)}\geq 1-\delta_{\operatorname{loss}}$.

Similarly, there exist $8$ primary interferers at a distance of at least $\sqrt{a_p}$, and $16$ primary interferers at a distance of at least $4\sqrt{a_p}$, and so on.
Then
\begin{equation} \label{EQ:upper_i_p}
I_p^i(n)=\sum_{k=1,k\neq
i}^{N_p(n)}P^k_pg\left(\|X_{p,\operatorname{tx}}^k-X_{p,\operatorname{rx}}^i\|\right)<P2^{\alpha/2+3}\sum_{t=1}^{\infty}t(3t-2)^{-\alpha}=I,
\end{equation}
where we use the fact that $P^k_p\leq P(\sqrt{2a_p})^{\alpha}$.
Thus,
\begin{equation}
R_{\operatorname{alone}}(n)>\frac{1}{9}\log\left(1+\frac{P}{N_0+I}\right)=K_p.
\end{equation}
Therefore, Lemma \ref{THM:primary_rate_adhoc1} holds.

\bigskip

\noindent{\it Proof of Lemma \ref{THM:primary_traffic_adhoc}}

Let $n_h$ denote the number of extended HDPs that should be delivered by a primary cell.
Similarly, $n_v$ denotes the number of extended VDPs that should be delivered by a primary cell.
When HDPs  are extended, the extended HDPs of all primary sources located in the area of ${1\times\sqrt{a_p}}$ should be handled by the primary cell.
By assuming that all primary nodes are sources, the resulting upper bound on $n_h$ follows $\mbox{Poisson}(\lambda=n\sqrt{a_p})$.
Using Lemma \ref{THM:upper_bound_poission}, we obtain
\begin{equation} \label{EQ:upper_n_ph}
\mathbb{P}(n_h\geq 2n\sqrt{a_p})\leq \frac{e^{-n\sqrt{a_p}}(e n\sqrt{a_p})^x}{x^x}\Big|_{x=2n\sqrt{a_p}}=e^{-n\sqrt{a_p}}\left(\frac{e}{2}\right)^{2n\sqrt{a_p}}.
\end{equation}
Similarly, the extended HDPs of all primary destinations located in the area of $\sqrt{a_p}\times1$ should be also handled by the primary cell.
By assuming that all primary nodes are destinations, we obtain
\begin{equation} \label{EQ:upper_n_pv}
\mathbb{P}(n_v\geq 2n\sqrt{a_p})\leq e^{-n\sqrt{a_p}}\left(\frac{e}{2}\right)^{2n\sqrt{a_p}}.
\end{equation}
From (\ref{EQ:upper_n_ph}) and (\ref{EQ:upper_n_pv}), we obtain
\begin{eqnarray}
\mathbb{P}(n_h+n_v\geq 4n\sqrt{a_p})\!\!\!\!\!\!\!\!\!&&\leq\mathbb{P}\left((n_h\geq 2n\sqrt{a_p})\cup (n_v\geq 2n\sqrt{a_p})\right)\nonumber\\
&&\leq 2e^{-n\sqrt{a_p}}\left(\frac{e}{2}\right)^{2n\sqrt{a_p}},
\end{eqnarray}
where the last inequality comes from the union bound.

Therefore, the probability that there is at least one primary cell supporting more than $4n\sqrt{a_p}$ extended data paths is upper bounded by $2ne^{-n\sqrt{a_p}}\left(\frac{e}{2}\right)^{2n\sqrt{a_p}}$, where the union bound and the fact that there are at most $n$ primary cells are used.
Since $2ne^{-n\sqrt{a_p}}\left(\frac{e}{2}\right)^{2n\sqrt{a_p}}\rightarrow0$ as $n\rightarrow\infty$, each primary cell should deliver the corresponding data of at most $4n\sqrt{a_p}$ extended data paths w.h.p., where $a_p=\frac{2\log n}{n}$.
Note that the above bounds also hold for the original data paths, which completes the proof.

%
%
%
\bigskip

\noindent{\it Proof of Lemma \ref{THM:number_of_served_SD_adhoc}}

Let $A_{p,1}$ denote the area of all preservation regions, $A_{p,2}$ denote the area of all disjoint regions due to the preservation regions except the biggest region, and $A_p=A_{p,1}+A_{p,2}$.
Define $m_p$ as the number of secondary nodes in the area of $A_p$ that follows $\mbox{Poisson}(\lambda=mA_p)$.
The number of secondary S-D pairs not served is clearly upper bounded by $m_p$.
From Lemma \ref{THM:upper_bound_poission}, we obtain
\begin{eqnarray} \label{EQ:upper_m_p}
\mathbb{P}(m_p\geq 2mA_p)=e^{-mA_p}\left(\frac{e}{2}\right)^{2mA_p}.
\end{eqnarray}
An upper bound on $A_{p,1}$ is obtained if we assume none of the regions overlap.
Thus, as each preservation region has an area of $9a_s$ and there are at most $(1+\epsilon)n$ such regions w.h.p., we obtain w.h.p.
\begin{equation} \label{EQ:upper_A_p_1}
A_{p,1}\leq9(1+\epsilon)na_s.
\end{equation}

\begin{figure}
  \begin{center}
  \scalebox{0.9}{\includegraphics{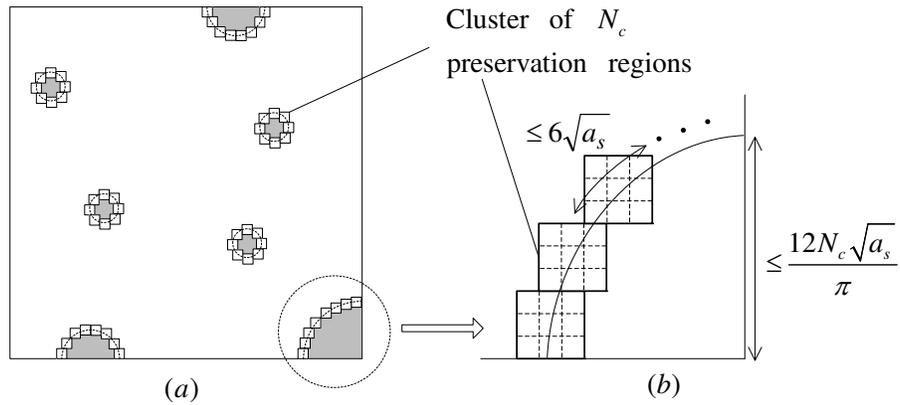}}
  \caption{Given that the size of any cluster of preservation regions is limited to $N_c$, this figure illustrates the worst-case scenario for the number of secondary S-D pairs that are not served when their data pathes are disconnected by the preservation regions.}
  \label{FIG:fraction}
  \end{center}
\end{figure}

To derive an upper bound on $A_{p,2}$, we assume all preservation regions form clusters having $N_c$ preservation region each (Corollary \ref{THM:clusters_of_preservation_regions}) shown in Fig. \ref{FIG:fraction}. (a), where the shaded regions denote $A_{p,2}$.
Then the maximum disjoint area generated by a cluster of $N_c$ preservation regions is given in Fig. \ref{FIG:fraction}. (b) as a circle maximizes the area of a region for a given perimeter.
Because each preservation region contributes a length of at most $6\sqrt{a_p}$ to the circumference of this circle, the radius is upper bounded by $\frac{12N_c\sqrt{a_s}}{\pi}$.
Thus, $A_{p,2}$ is upper bounded w.h.p. by
\begin{eqnarray} \label{EQ:upper_A_p_2}
A_{p,2}<\frac{(1+\epsilon)n}{N_c}\frac{\pi}{4}\left(\frac{12N_c\sqrt{a_s}}{\pi}\right)^2=\frac{36N_c(1+\epsilon)}{\pi}na_s,
\end{eqnarray}
where we use the fact that the total number of clusters having $N_c$ preservation regions in each cluster is upper bounded by $\frac{(1+\epsilon)n}{N_c}$ w.h.p..
From (\ref{EQ:upper_A_p_1}) and (\ref{EQ:upper_A_p_2}), $A_p$ is upper bounded by $18\beta(1+\epsilon)\frac{\pi+4N_c}{\pi}n^{1-\beta}\log n$ w.h.p..
By substituting $A_p$ for its upper bound in ($\ref{EQ:upper_m_p}$), we obtain
\begin{eqnarray} \label{EQ:upper_m_p2}
&&\mathbb{P}\left(m_p\geq 36\beta(1+\epsilon)\frac{\pi+4N_c}{\pi}n\log n\right)\nonumber\\
&&\leq e^{-18\beta(1+\epsilon)\frac{\pi+4N_c}{\pi}n\log n}\left(\frac{e}{2}\right)^{36\beta(1+\epsilon)\frac{\pi+4N_c}{\pi}n\log n}\rightarrow0\mbox{ as }n\rightarrow\infty.
\end{eqnarray}
Thus, we obtain w.h.p.
\begin{equation}
m_p<\epsilon_{s,1}(m)(1-\epsilon)\frac{m}{2},
\end{equation}
where $\epsilon_{s,1}(m)=72\frac{1+\epsilon}{1-\epsilon}\frac{\pi+4N_c}{\pi}\frac{\log m}{m^{1-1/\beta}}$.
Since the total number of secondary S-D pairs is lower bounded by $(1-\epsilon)\frac{m}{2}$ w.h.p., the fraction of unserved S-D pairs is upper bounded by $\epsilon_{s,1}(m)$ w.h.p., which completes the proof.

\bigskip

\noindent{\it Proof of Lemma \ref{THM:secondary_rate_adhoc}}

Since the same secondary packet is transmitted three times, the minimum distance of $\frac{\sqrt{a_p}}{2}$ from all primary interferers to the secondary Rx can be guaranteed for one out of three transmissions.
Then the interference from primary interferers of that packet is upper bounded by
\begin{eqnarray}
I_{ps}\!\!\!\!\!\!\!\!\!&&<P(\sqrt{2a_p})^{\alpha}\sum_{t=1}^{\infty}8k((3t-2)\sqrt{a_p})^{-\alpha}+P(\sqrt{2a_p})^{\alpha}\left(\frac{\sqrt{a_p}}{2}\right)^{-\alpha}\nonumber\\
&&=I+2^{3\alpha/2}P,
\end{eqnarray}
where we use the same technique as in Lemma \ref{THM:primary_rate_adhoc1}.
Similarly, $I_{s}$ is lower bounded by $\delta_P I$.
Thus, the rate of each secondary cell is lower bounded by
\begin{equation} \label{EQ:lower_secondary_rate_adhoc}
\frac{1}{27}\log\left(1+\frac{\delta_{P}P}{N_0+(1+\delta_P)I+2^{3\alpha/2}P}\right)=K_s,
\end{equation}
where $\frac{1}{27}$ indicates the rate loss due to the $9$-TDMA and repeated (three times) transmissions of the same secondary packet.
Therefore, Lemma \ref{THM:secondary_rate_adhoc} holds.

\bigskip

\noindent{\it Proof of Lemma \ref{THM:secondary_traffic_adhoc}}

Let $m_{h,1}$ and $m_{h,2}$ denote the number of extended HDPs including re-routed paths that should be delivered by a secondary regular cell and by a secondary loaded cell, respectively.
Similarly, we can define $m_{v,1}$ and $m_{v,2}$ for extended VDPs.

Let us first consider a regular cell.
This regular cell delivers the corresponding data of extended HDPs passing through it.
Then all extended HDPs of the secondary sources located in the area of $1\times\sqrt{a_s}$ should be handled by the regular cell, where we ignore the effect of S-D pairs not served, which yields an upper bound on the total number of HDPs.
By assuming that all secondary nodes are sources, the resulting upper bound on $m_{h,1}$ follows $\mbox{Poisson}(\lambda=m\sqrt{a_s})$.
From Lemma \ref{THM:upper_bound_poission}, we obtain
\begin{equation} \label{EQ:upper_m_sh_1}
\mathbb{P}(m_{h,1}\geq 2m\sqrt{a_s})\leq e^{-m\sqrt{a_s}}\left(\frac{e}{2}\right)^{2m\sqrt{a_s}}.
\end{equation}
We obtain the same bound for $m_{v,1}$ by assuming that all secondary nodes are destinations and then
\begin{eqnarray} \label{EQ:upper_m_sh_1_sv_1}
\mathbb{P}(m_{h,1}+m_{v,1}\geq 4m\sqrt{a_s})\!\!\!\!\!\!\!\!\!&&\leq\mathbb{P}\left((m_{h,1}\geq 2m\sqrt{a_s})\cup(m_{v,2}\geq 2m\sqrt{a_s})\right)\nonumber\\
&&\leq 2e^{-m\sqrt{a_s}}\left(\frac{e}{2}\right)^{2m\sqrt{a_s}}.
\end{eqnarray}
From the union bound and the fact that there are at most $m$ secondary cells, each regular cell should deliver the corresponding data of at most $4m\sqrt{a_s}$ extended data paths w.h.p., where we use the fact that $2me^{-m\sqrt{a_s}}\left(\frac{e}{2}\right)^{2m\sqrt{a_s}}\rightarrow0$ as $m\rightarrow\infty$.

\begin{figure}
  \begin{center}
  \scalebox{1.1}{\includegraphics{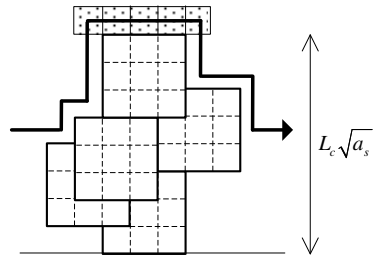}}
  \caption{An upper bound on the number of re-routed HDPs passing through the dotted cells.}
  \label{FIG:cluster_preservation_region}
  \end{center}
\end{figure}

Let us now consider a loaded cell.
Unlike in the primary data path which has no obstacles, a secondary data path should circumvent any preservation regions which lie on its path.
Therefore, the loaded cells should deliver more data paths than the regular cells w.h.p..
Suppose a cluster of preservation regions located on the boundary of the network in Fig. \ref{FIG:cluster_preservation_region}, whose projection on $y$-axis has a length of $L_c\sqrt{a_s}$.
Then all extended HDPs of the secondary sources located in the area of $1\times L_c\sqrt{a_s}$ is re-routed through the dotted cells, where we ignore the effect of S-D pairs not served (which yields an upper bound on the total number of extended HDPs).
The other loaded cells will deliver less HDPs than the dotted cells w.h.p..
Recall that $L_c\leq3N_c$ w.h.p. (Corollary \ref{THM:clusters_of_preservation_regions}) and the dotted cells need to deliver re-routing paths of at most two such clusters.
Therefore, by assuming that all secondary nodes are sources, the resulting upper bound on $m_{h,2}$ follows $\mbox{Poisson}(\lambda=m(6N_c+1)\sqrt{a_s})$.
Note that the upper bound on $m_{h,2}$ is the same as the upper bound on $m_{h,1}$ except for a constant factor of $6N_c+1$, where $6N_c$ comes from the re-routed HDPs of two adjacent clusters and $1$ comes from the original HDPs.
Therefore, we can apply the same analysis used in the regular case.
In conclusion, each loaded cell should deliver the corresponding data of at most $4m(6N_c+1)\sqrt{a_s}$ extended data paths w.h.p..
Since the above bounds also hold for the original data paths, Lemma \ref{THM:secondary_traffic_adhoc} holds.

\bigskip
\noindent{\it Proof of Lemma \ref{THM:primary_rate_infra}}

The overall procedure of the proof is similar to that of Lemma \ref{THM:primary_rate_adhoc1}.
Let us first consider downlink transmissions, where all primary cells are activated simultaneously at a given moment.
Let $I'_{p,d}$ and $I'_{sp,d}$ denote the interference from all primary interferers and all secondary interferers during downlink, respectively.
Let $R'_{p,d}$ and $R'_{\operatorname{alone},d}$ denote the downlink rates of a primary cell in the presence of the secondary network and in the absence of the secondary network, respectively.
Then $R'_{\operatorname{alone},d}=R'_{p,d}$ if $I'_{sp,d}=0$.
From the same bounds in (\ref{EQ:upper_i_sp}) and (\ref{EQ:fraction_lower}), we obtain $\lim_{l\to\infty}\frac{R'_{p,d}}{R'_{\operatorname{alone},d}}\geq 1-\delta_{\operatorname{loss}}$ for $\delta_P\in(0, \min\{\delta'_{P, \operatorname{max}},1\})$.
The same bound can be derived for the uplink.
Thus, (\ref{EQ:R_p_R_alone}) holds.

\begin{figure}
  \begin{center}
  \scalebox{0.65}{\includegraphics{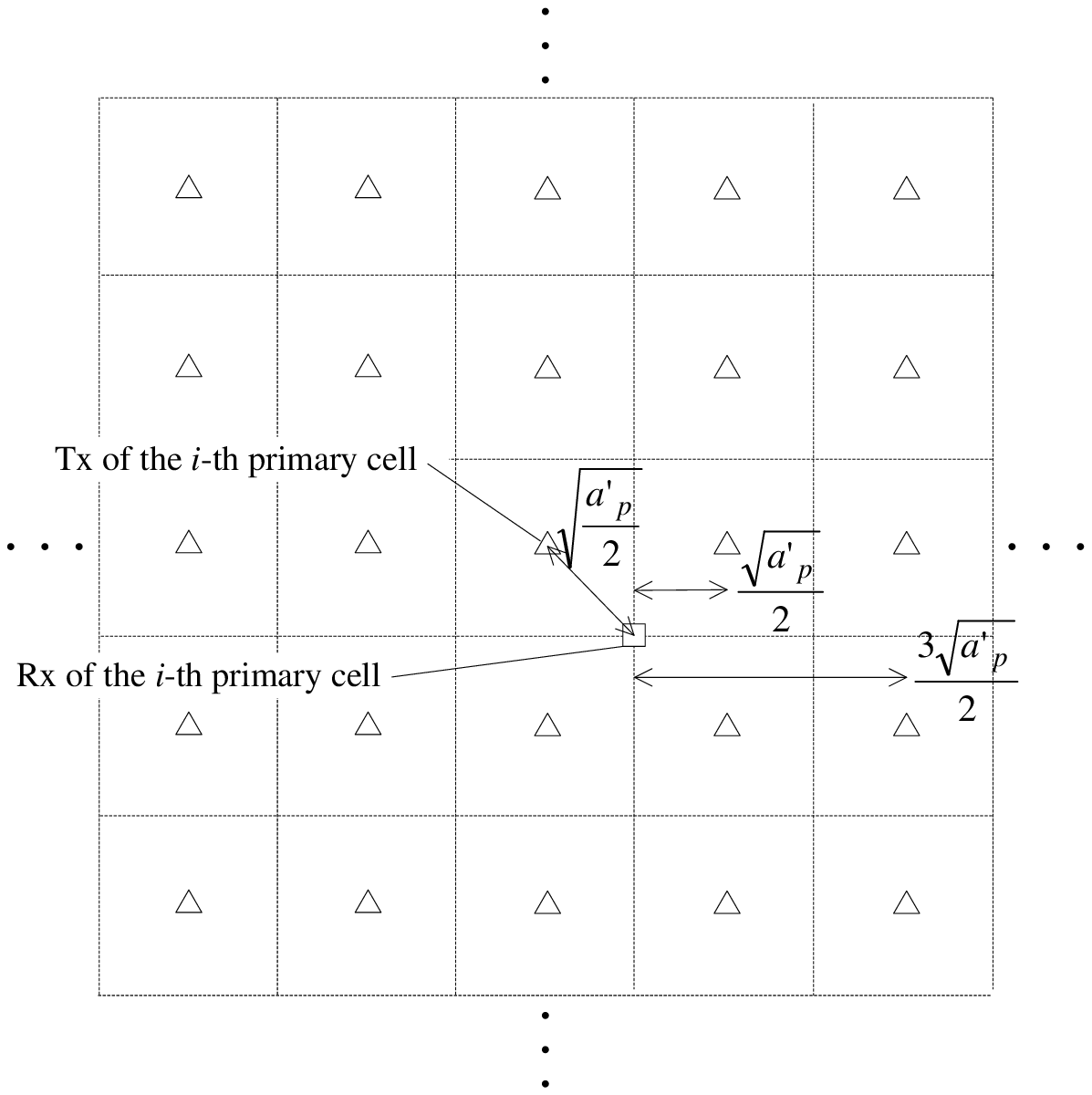}}
  \caption{The amount of interference from the primary interferers to the Rx of the $i$-th active primary cell for the infrastructure-supported primary model during downlink transmissions.}
  \label{FIG:i_p_infra}
  \end{center}
\end{figure}

Now consider the bound on $I'_{p,d}$.
Since there exist $8$ primary interferers at a distance of at least $\frac{1}{2}\sqrt{a'_p}$ and $16$ primary interferers at a distance of at least $\frac{3}{2}\sqrt{a'_p}$ and so on (see Fig. \ref{FIG:i_p_infra}), we obtain
\begin{equation}
I'_{p,d}<P\left(\sqrt{a'_p/2}\right)^{\alpha}\sum_{t=1}^{\infty}8t\left((2t-1)\frac{\sqrt{a'_p}}{2}\right)^{-\alpha}=I',
\end{equation}
where we use the fact that the transmit power of each BS is upper bounded by $P\left(\sqrt{a'_p/2}\right)^{\alpha}$.
Then
\begin{equation} \label{EQ:lower_r_pd}
R'_{\operatorname{alone},d}>\log\left(1+\frac{P}{N_0+I'}\right)=K'_p.
\end{equation}
In a similar manner, the rate of each primary cell during uplink is also lower bounded by $K'_p$.
Therefore, we can guarantee a constant rate of $K'_p$ for each primary cell during both downlink and uplink, which completes the proof.

\bigskip

\noindent{\it Proof of Lemma \ref{THM:primary_traffic_infra}}

Let $n'_p$ denote the number of primary nodes in a primary cell, which follows $\mbox{Poisson}\left(\lambda=na'_p\right)$.
From Lemma \ref{THM:upper_bound_poission}, we obtain
\begin{equation}
\mathbb{P}(n'_p\geq 2na'_p)\leq e^{-na'_p}\left(\frac{e}{2}\right)^{2na'_p}.
\end{equation}
From the union bound, each primary cell has at most $2na'_p$ primary nodes w.h.p., where we use the fact that $ne^{-na'_p}\left(\frac{e}{2}\right)^{2na'_p}\rightarrow 0$ as $n\rightarrow\infty$.
If we assume that all primary nodes are destinations (or sources), the number of downlink transmissions (or the number of uplink transmissions) per primary cell is upper bounded by $2na'_p=2n^{1-\gamma}$ w.h.p..
Therefore, the lemma holds.

\bigskip

\noindent{\it Proof of Lemma \ref{THM:number_of_served_SD_infra}}

Let $A_b$ denote the area of all preservation regions around BSs and $m_b$ denote the number of secondary nodes in the area of $A_b$.
Then, From Lemma \ref{THM:upper_bound_poission},
\begin{equation}
\mathbb{P}(m_b\geq 2mA_b)\leq e^{-mA_b}\left(\frac{e}{2}\right)^{2mA_b}.
\end{equation}
Since each preservation region around BS has an area of $\frac{\delta_a a'_p}{\log n}$ and there are $l$ such regions, which are not overlapping with each other, $A_b=\frac{\delta_a}{\log n}$.
Thus, we know $m_b< \epsilon_b(m)(1-\epsilon)\frac{m}{2}$ w.h.p., where
\begin{equation}
\epsilon_b(m)=\frac{4\beta\delta_a}{(1-\epsilon)\log m}=\Theta\left(\frac{1}{\log m}\right).
\end{equation}
Combining this with the result of Lemma \ref{THM:number_of_served_SD_adhoc}, we obtain $m_p+m_b<(\epsilon_{s,1}(m)+\epsilon_b(m))(1-\epsilon)\frac{m}{2}$ w.h.p..
Since the number of S-D pairs not served is clearly upper bounded by $m_p+m_b$, the fraction of unserved S-D pairs is upper bounded by $\epsilon'_{s,1}(m)=\epsilon_{s,1}(m)+\epsilon_b(m)=\Theta(\frac{1}{\log m})$ w.h.p., which completes the proof.

\bigskip

\noindent{\it Proof of Lemma \ref{THM:secondary_rate_infra}}

First consider the rate of a secondary cell in the avoidance regions (but not in the preservation regions).
Due to the preservation regions around BSs, the minimum distance of $\frac{1}{2}\sqrt{\frac{\delta_a}{\log n}a'_p}$ can be guaranteed from all primary interferers.
Thus, $I'_{ps}<I'+\left(\frac{1}{2}\sqrt{\frac{\delta_a}{\log n}a'_p}\right)^{-\alpha}P\left(\sqrt{\frac{1}{2}a'_p}\right)^{\alpha}=I'+P(\frac{2\log m}{\beta\delta_a})^{\alpha/2}$.
Similarly $I'_{s}<\delta_PI$.
Then the rate of each secondary cell in the avoidance regions is upper bounded by
\begin{equation}
\frac{\delta_t}{18}\log\left(1+\frac{\delta_P P}{N_0+I'+\delta_P I+P\left(\frac{2\log m}{\beta\delta_a}\right)^{\alpha/2}}\right)=K'_{s,1}(m),
\end{equation}
where $\frac{\delta}{18}$ arises from $9$-TDMA, the time fraction of Phase 1, and the time fraction of downlink.

In the case of a secondary cell outside the avoidance regions, the minimum distance of $\frac{1}{2}\sqrt{\delta_a a'_p}$ can be guaranteed from all primary interferers.
Then the rate of each secondary cell outside the avoidance regions is upper bounded by
\begin{equation}
\frac{1-\delta_t}{18}\log\left(1+\frac{\delta_PP}{N_0+I'+\delta_P I+P\left(\frac{2}{\delta_a}\right)^{\alpha/2}}\right)=K'_{s,2},
\end{equation}
where $\frac{1-\delta}{18}$ arises from $9$-TDMA, the time fraction of Phase 2, and the time fraction of downlink.
Therefore, Lemma \ref{THM:secondary_rate_infra} holds.

\bigskip

\noindent{\it Proof of Lemma \ref{THM:secondary_traffic_infra}}

Consider Phase $2$ in which the secondary cells outside the avoidance regions are activated.
Let $m'_{h,1}$ and $m'_{h,2}$ denote the number of extended HDPs that should be delivered by a secondary regular cell and by a secondary loaded cell, respectively.
We can define $m'_{v,1}$ and $m'_{v,2}$ analogously for VDPs.

\begin{figure}
  \begin{center}
  \scalebox{0.7}{\includegraphics{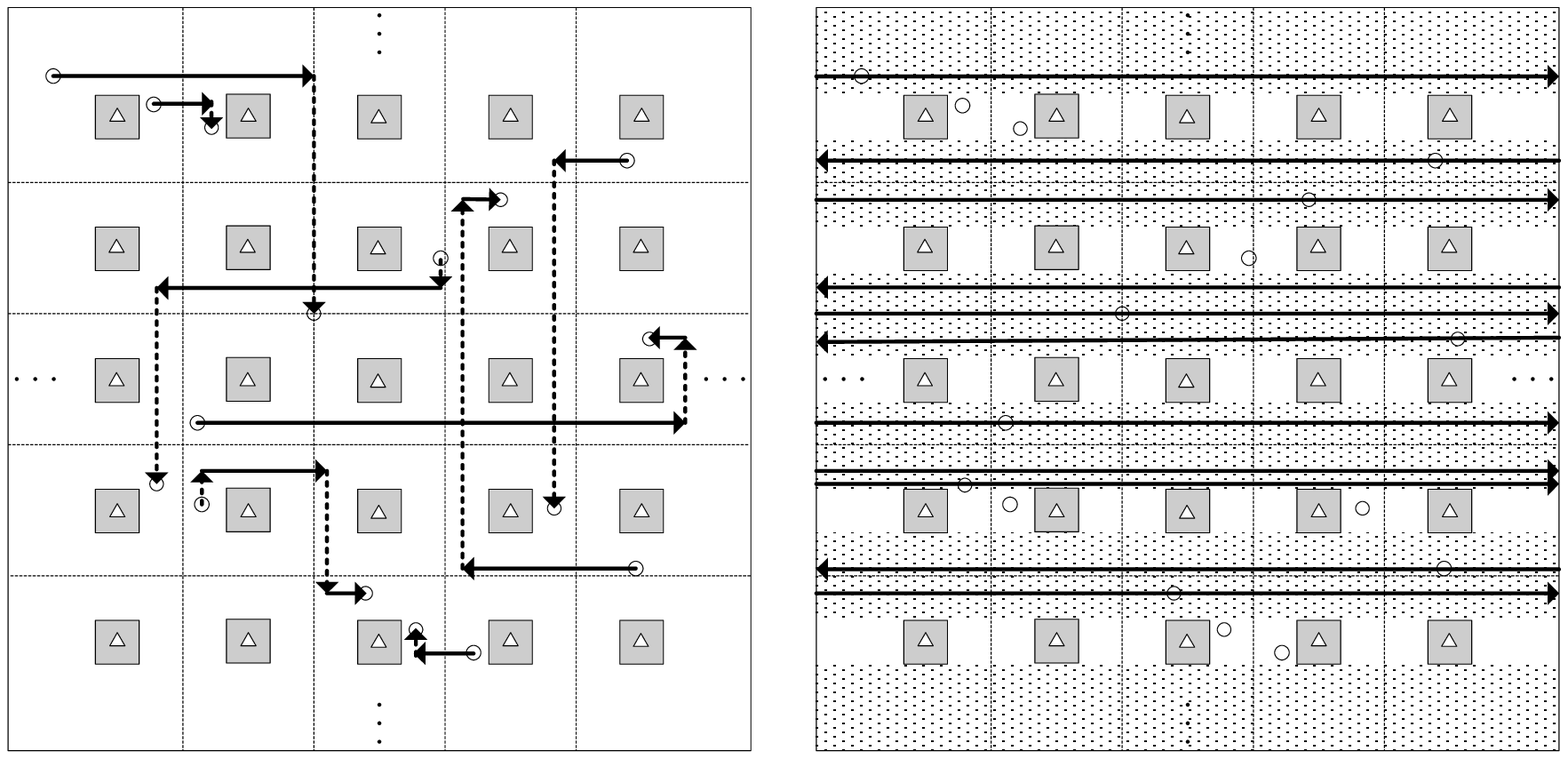}}
  \caption{Examples of original (or shifted) HDPs (left) and their extended HDPs (right) in $\mathcal{R}_h$ of the secondary S-D pairs for the infrastructure-supported primary model, where the dotted regions are denoted by $\mathcal{R}_h$. For simplicity, the preservation regions are not shown in this figure.}
  \label{FIG:routing_analysis_infra}
  \end{center}
\end{figure}

Let us first consider a regular cell in $\mathcal{R}_h\cap\mathcal{R}_v$.
There are two types of HDPs in $\mathcal{R}_h$: the first type is an original (or a shifted) HDP and the second type is a short horizontal hops in order to reach each destination.
Note that a short HDP only occurs if its original VDP is blocked by an avoidance region.
We assume that a short HDP always occurs regardless of its VDP and extend it to the entire horizontal line including the short HDP.
Fig. \ref{FIG:routing_analysis_infra} illustrates examples of original (or shifted) HDPs (left) and their extended HDPs (right) in $\mathcal{R}_h$.
Note that the $y$-axis of an extended HDP from an original (or shifted) HDP originates from a source node.
Similarly, the $y$-axis of an extended HDP from a short HDP originates from a destination node.
As a result, under this extended traffic, all secondary nodes generate extended HDPs on $\mathcal{R}_h$ because each node is a source or a destination, where we ignore the effects of the S-D pairs not served and the S-D pairs that do not generate traffic on $\mathcal{R}_h$.
Since a regular cell in $\mathcal{R}_h$ delivers the corresponding data of all extended HDPs passing through it, all extended HDPs of the secondary nodes located in the area of $1\times\sqrt{a'_s}$ should be delivered by the regular cell.
Additionally, it should deliver the corresponding data of all nodes in the area of $1\times\frac{D_1}{D_2}\sqrt{a'_s}$ because these extended HDPs are shifted to $\mathcal{R}_h$.
Therefore, the resulting upper bound on $m'_{h,1}$ follows $\mbox{Poisson}\left(\lambda=m\frac{D_1+D_2}{D_2}\sqrt{a'_s}=mc\sqrt{a'_s}\right)$, where $c=(1-\sqrt{\delta_a})^{-1}$.
From Lemma \ref{THM:upper_bound_poission}, we obtain
\begin{eqnarray}
\mathbb{P}\left(m'_{h,1}\geq 2mc\sqrt{a'_s}\right)\leq e^{-mc\sqrt{a'_s}}\left(\frac{e}{2}\right)^{2mc\sqrt{a'_s}}.
\end{eqnarray}
The same bound can be obtained for $m'_{v,1}$.
From the fact that the number of data paths that should be delivered by a regular cell in $\mathcal{R}_h\cap\mathcal{R}_v$ is given by $m'_{h,1}+m'_{v,1}$, we obtain
\begin{eqnarray}
\mathbb{P}\left(m'_{h,1}+m'_{v,1}\geq 4mc\sqrt{a'_s}\right)\!\!\!\!\!\!\!\!\!&&\leq\mathbb{P}\left(\left(m'_{s,h,1}\geq 2mc\sqrt{a'_s}\right)\cup\left(m'_{s,v,1}\geq 2mc\sqrt{a'_s}\right)\right)\nonumber\\
&&\leq 2e^{-mc\sqrt{a'_s}}\left(\frac{e}{2}\right)^{2mc\sqrt{a'_s}}.
\end{eqnarray}
By the union bound and the fact that there are at most $m$ secondary cells, each regular cell in $\mathcal{R}_h\cap\mathcal{R}_v$ should deliver at most $4mc\sqrt{a'_s}$ extended data paths w.h.p., where we use the fact $2me^{-mc\sqrt{a'_s}}\left(\frac{e}{2}\right)^{2mc\sqrt{a'_s}}\rightarrow 0$ as $n\rightarrow\infty$.

Unlike the previous case, all S-D pairs that generate HDPs in $\mathcal{R}^c_h$ are not vertically blocked such that only original HDPs exist in $\mathcal{R}^c_h$.
Then, $m'_{h,1}$ is upper bounded by $2m\sqrt{a'_s}$ w.h.p. in this case.
Therefore the regular cells in $\mathcal{R}^c_h\cap\mathcal{R}_v$, $\mathcal{R}_h\cap\mathcal{R}^c_v$, and $\mathcal{R}^c_h\cap\mathcal{R}^c_v$ deliver w.h.p. less data paths compared to the regular cells in $\mathcal{R}_h\cap\mathcal{R}_v$.
In conclusion, each regular cell should deliver the corresponding data of at most $4c\sqrt{2m\log m}$ extended data paths w.h.p..

To obtain an upper bound on $m'_{h,2}$, consider again the cluster of the preservation regions located on the boundary of the network in Fig. \ref{FIG:cluster_preservation_region} (or the boundary of an avoidance region in this case).
Then all nodes located in the area of $1\times (2L_c+1)\sqrt{a'_s}$ generate extended HDPs passing through the dotted cells in $\mathcal{R}_h$.
Additionally, all nodes located in the area of $1\times\frac{D_1}{D_2}\left(2L_c+1\right)\sqrt{a'_s}$, belonging to $\mathcal{R}^c_h$, generate extended HDPs passing through the dotted cells since they are shifted to $\mathcal{R}_h$.
Thus, from the fact $L_c\leq 3N_c$ w.h.p., $m'_{h,2}\leq 2(6N_c+1)c\sqrt{2m\log m}$ w.h.p..
By applying the same bound on $m'_{v,2}$, we conclude that each loaded cell should deliver the corresponding data of at most $4(6N_c+1)c\sqrt{2m\log m}$ data paths w.h.p..
Note that the loaded cells in $\mathcal{R}^c_h\cap\mathcal{R}_v$, $\mathcal{R}_h\cap\mathcal{R}^c_v$, and $\mathcal{R}^c_h\cap\mathcal{R}^c_v$ deliver w.h.p. less data paths than the loaded cells in $\mathcal{R}_h\cap\mathcal{R}_v$.
Thus, Lemma \ref{THM:secondary_traffic_infra} holds.

\bigskip

\noindent{\it Proof of Lemma \ref{THM:secondary_traffic_infra2}}

Consider Phase $1$ in which the secondary cells in the avoidance regions are activated.
Since the avoidance regions are in $\mathcal{R}^c_h\cup\mathcal{R}^c_v$, there exists no shifted data path.
The overall procedure is similar to the proof of Lemma \ref{THM:secondary_traffic_infra}.
Let us first consider the secondary regular cells.
If we extend HDP to the line having the length of $\frac{1}{2}\sqrt{\delta_a a'_p}$, which is the length of half an avoidance region side, all nodes in the area of $\frac{1}{2}\sqrt{\delta_a a'_p}\times \sqrt{a'_s}$ generate extended HDPs passing through a regular cell.
Thus, the number of extended HDPs delivered by each regular cell is upper bounded by $\sqrt{\delta_a a'_p}\times \sqrt{a'_s}m=\sqrt{2\delta_a m^{1-\gamma/\beta}\log m }$ w.h.p..
By the same analysis for VDP, each regular cell should deliver the corresponding data of at most $2\sqrt{2\delta_a m^{1-\gamma/\beta}\log m }$ extended data paths w.h.p..
Similarly, each secondary loaded cell should deliver the corresponding data of at most $2\left(6N_c+1\right)\sqrt{2\delta_a m^{1-\gamma/\beta}\log m }$ extended data paths w.h.p., which completes the proof.


\newpage

\begin{table}
\caption{Definition of symbols related to achievable rates for each primary and secondary transmit pair.}
\label{Table:simbols}
\begin{equation*}
  \begin{array}{|c|c|}
  \hline
  P_p^i & \text{Transmit power of the $i$-th primary pair}\\
  \hline
  P_s^j & \text{Transmit power of the $j$-th secondary pair}\\
  \hline
  N_0 & \text{Thermal noise power}\\
  \hline
  X_{p,\text{tx}}^i& \text{Tx location of the $i$-th primary pair}\\
  \hline
  X_{p,\text{rx}}^i& \text{Rx location of the $i$-th primary pair}\\
  \hline
  X_{s,\text{tx}}^j& \text{Tx location of the $j$-th secondary pair}\\
  \hline
  X_{s,\text{rx}}^j& \text{Rx location of the $j$-th secondary pair}\\
  \hline
  I_p^i&\text{Interference power from the primary Txs to the Rx of the $i$-th primary pair}\\
  \hline
  I_{sp}^i&\text{Interference power from the secondary Txs to the Rx of the $i$-th primary pair}\\
  \hline
  I_s^j&\text{Interference power from the secondary Txs to the Rx of the $j$-th secondary pair}\\
  \hline
  I_{ps}^j&\text{Interference power from the primary Txs to the Rx of the $j$-th secondary pair}\\
  \hline
  R_p^i&\text{Rate of the $i$-th primary pair}\\
  \hline
  R_s^j&\text{Rate of the $j$-th secondary pair}\\
  \hline
  \end{array}
\end{equation*}
\end{table}

\end{document}